\documentclass
	[
		10 pt
	]
	{article}

\usepackage
	[
		a4paper,
		left = 2.500 cm,
		right = 2.500 cm,
		top = 2.500 cm,
		bottom = 3.000 cm
	]
	{geometry}
\usepackage{tikz, pgf, pgfplots}
\usetikzlibrary
	{
		3d,
		arrows,											
		arrows.meta,
		automata,										
		backgrounds,
		bending,
		calc,
		chains,
		datavisualization.formats.functions,			
		decorations.pathmorphing,
		decorations.pathreplacing,
		decorations.text,
		fadings,
		fit,
		graphs,
		graphs.standard,
		matrix,
		positioning,									
		quantikz2,
		quotes,
		shadings,
		shadows,
		shadows.blur,
		shapes,
		shapes.geometric,
		trees
	}
\pgfplotsset{compat = newest}
\usepgflibrary
	{
		shapes.misc
	}
\usepgfplotslibrary
	{
		dateplot
	}
\usepackage{tikzpeople}
\usepackage{fontawesome5}
\usepackage{smartdiagram}

\allowdisplaybreaks
\usepackage{amsthm}
\usepackage{amssymb}
\usepackage[bb = boondox]{mathalfa}
\usepackage{dsfont}
\usepackage{newtxmath}
\usepackage{mathtools}
\usepackage{multicol}
\usepackage{physics2}
\usephysicsmodule{braket}
\usepackage{empheq}
\usepackage[ compat = newest ]{yquant}

\usepackage{sectsty}
\allsectionsfont{\color{WordBlue}}
\usepackage{xcolor}
\usepackage{varwidth}
\usepackage{graphicx}
\usepackage{wrapfig}
\usepackage{xurl}                                                                                  %
\usepackage{hyperref}
\hypersetup
	{
		colorlinks,
		linkcolor = {WordRed!75!black},
		citecolor = {WordBlueDarker25},
		urlcolor = {WordAquaDarker50}
	}
\usepackage{tabularray}
\usepackage{colortbl}
\usepackage{float}
\usepackage{nicematrix}
\usepackage{cellspace}
\usepackage{makecell}
\usepackage{multirow}
\usepackage{caption}
\usepackage[ linesnumbered, tworuled, vlined ] {algorithm2e}
\usepackage{appendix}
\usepackage{enumitem}
\usepackage{siunitx}
\usepackage{xintexpr}
\usepackage{spreadtab}
\usepackage{framed}
\usepackage{svg}
\usepackage{lipsum}
\definecolor{MyLightRed}{RGB}{244, 213, 245}
\definecolor{WordRed}{RGB}{255, 0, 102}
\definecolor{WordRedAccent5Lighter60}{HTML}{F5B5A7}
\definecolor{WordRedAccent5Darker25}{HTML}{B23214}
\definecolor{RedDarkLightest}{HTML}{ff0088}
\definecolor{RedDarkLight}{HTML}{ea005f}
\definecolor{RedPurple}{HTML}{aa007f}
\definecolor{Purple}{HTML}{911146}
\definecolor{PurpleDark}{RGB}{102, 0, 102}
\definecolor{WordLightGreen}{RGB}{140, 214, 192}
\definecolor{WordGreen}{RGB}{0, 176, 80}
\definecolor{GreenLightest}{HTML}{00ffa0}
\definecolor{GreenLighter1}{HTML}{00b383}
\definecolor{GreenLighter2}{HTML}{00aa7f}
\definecolor{GreenDark}{HTML}{225522}
\definecolor{GreenTeal}{HTML}{008080}
\definecolor{WordIceBlue}{RGB}{223, 227, 229}
\definecolor{MyVeryLightBlue}{RGB}{211, 245, 247}
\definecolor{WordBlueVeryLight}{RGB}{0, 176, 240}
\definecolor{WordBlueLight}{RGB}{0, 112, 192}
\definecolor{WordBlueDark}{RGB}{46, 116, 181}
\definecolor{WordBlueDarker}{RGB}{31, 78, 121}
\definecolor{WordBlueDarker25}{RGB}{54, 96, 146}
\definecolor{WordBlueDarker50}{RGB}{36, 64, 98}
\definecolor{WordBlueDarkest}{RGB}{0, 32, 96}
\definecolor{WordBlue}{RGB}{19, 65, 99}
\definecolor{MyBlue}{RGB}{0, 64, 128}
\definecolor{MyDarkBlue}{RGB}{0, 51, 102}
\definecolor{BlueVeryDark}{HTML}{222255}
\definecolor{MagentaVeryLight}{RGB}{178, 162, 201}
\definecolor{MagentaLighter}{RGB}{161, 106, 221}
\definecolor{MagentaLight}{RGB}{128, 100, 162}
\definecolor{MagentaDark}{RGB}{106, 65, 152}
\definecolor{MagentaVeryDark}{RGB}{97, 75, 128}
\definecolor{WordAquaLighter80}{RGB}{218, 238, 243}
\definecolor{WordAquaLighter60}{RGB}{183, 222, 232}
\definecolor{WordAquaLighter40}{RGB}{146, 205, 220}
\definecolor{WordAquaDarker25}{RGB}{49, 134, 155}
\definecolor{WordAquaAccent2Darker25}{HTML}{398E98}
\definecolor{WordAquaDarker50}{RGB}{33, 89, 103}
\definecolor{WordVeryLightTeal}{RGB}{223, 236, 235}
\definecolor{WordLightTeal}{RGB}{160, 199, 197}
\definecolor{WordDarkTealLighter80}{RGB}{207, 223, 234}
\definecolor{WordDarkTeal}{RGB}{72, 123, 119}
\definecolor{WordDarkerTeal}{RGB}{48, 82, 80}
\definecolor{WordTurquoiseLighter80}{RGB}{209, 238, 249}
\definecolor{WordGoldAccent1Lighter40}{HTML}{FFDF6A}
\definecolor{WordGoldAccent1Darker25}{HTML}{C49A00}
\definecolor{Brown}{HTML}{666633}
\definecolor{WordOrangeAccent2Lighter60}{HTML}{FCD3A4}
\definecolor{WordOrangeAccent4Lighter60}{HTML}{F7C5A1}
\definecolor{LavenderBlush}{RGB}{255, 240, 245}
\definecolor{MediumTurquoise}{RGB}{72, 209, 204}
\definecolor{PowderBlue}{RGB}{176, 224, 230}
\definecolor{SkyBlue}{RGB}{135, 206, 235}
\definecolor{Azure2}{RGB}{224, 238, 238}
\definecolor{Azure3}{RGB}{193, 205, 205}
\definecolor{CadetBlue4}{RGB}{83, 134, 139}
\definecolor{DarkSeaGreen1}{RGB}{193, 255, 193}
\definecolor{DeepPink4}{RGB}{139, 10, 80}
\definecolor{Honeydew2}{RGB}{224, 238, 224}
\definecolor{LightSkyBlue1}{RGB}{176, 226, 255}
\definecolor{LightSkyBlue3}{RGB}{141, 182, 205}
\definecolor{LightSkyBlue4}{RGB}{96, 123, 139}
\definecolor{LightSteelBlue1}{RGB}{202, 225, 255}
\definecolor{LightSteelBlue4}{RGB}{110, 123, 139}
\definecolor{MediumPurple1}{RGB}{171, 130, 255}
\definecolor{PaleTurquoise3}{RGB}{150, 205, 205}
\definecolor{PaleVioletRed3}{RGB}{205, 104, 137}
\definecolor{Purple1}{RGB}{155, 48, 255}
\definecolor{SeaGreen1}{RGB}{84, 255, 159}
\definecolor{SeaGreen2}{RGB}{78, 238, 148}
\definecolor{SeaGreen3}{RGB}{67, 205, 128}
\definecolor{SkyBlue4}{RGB}{74, 112, 139}
\definecolor{SteelBlue1}{RGB}{99, 184, 255}
\definecolor{Thistle3}{RGB}{205, 181, 205}
\definecolor{Turquoise4}{RGB}{0, 134, 139}
\definecolor{VioletRed1}{RGB}{255, 62, 150}
\definecolor{VioletRed2}{RGB}{208, 32, 144}
\definecolor{VioletRed3}{RGB}{199, 21, 133}
\definecolor{VioletRed4}{RGB}{139, 10, 80}
\usepackage
	[
		skins,
		breakable,
		listings,
		raster,
		theorems
	]
	{tcolorbox}
\NewTcbTheorem
	[
		number within = section
	]
	{definition}
	{Definition}
	{
		breakable,
		skin = enhanced,
		enhanced jigsaw,			
		attach boxed title to top left = { xshift = -5.000 mm, yshift = 0.000 mm },
		boxed title style = { boxrule = 0.000 pt, sharp corners = all },
		varwidth boxed title,
		fonttitle = \bfseries,
		coltitle = SkyBlue!30!black,
		colbacktitle = SkyBlue!70!white,
		colframe = SkyBlue,
		colback = SkyBlue!20!white,
		sharp corners = all,
		toprule = 0.000 mm,
		bottomrule = 0.500 mm,
		leftrule = 0.500 mm,
		rightrule = 0.000 mm,
	}
	{def}
\NewTcbTheorem
	[
		number within = section
	]
	{example}
	{Example}
	{
		breakable,
		skin = enhanced,
		enhanced jigsaw,			
		attach boxed title to top left = { xshift = -5.000 mm, yshift = 0.000 mm },
		boxed title style = { boxrule = 0.000 pt, sharp corners = all },
		varwidth boxed title,
		fonttitle = \bfseries,
		coltitle = PowderBlue!30!black,
		colbacktitle = PowderBlue!70!white,
		colframe = PowderBlue,
		colback = white,
		sharp corners = all,
		toprule = 0.000 mm,
		bottomrule = 0.500 mm,
		leftrule = 0.500 mm,
		rightrule = 0.000 mm,
	}
	{xmp}
\NewTcbTheorem
	[
	number within = section
	]
	{theorem}
	{Theorem}
	{
		breakable,
		skin = enhanced,
		enhanced jigsaw,			
		attach boxed title to top left = { xshift = -5.000 mm, yshift = 0.000 mm },
		boxed title style = { boxrule = 0.000 pt, sharp corners = all },
		varwidth boxed title,
		fonttitle = \bfseries,
		coltitle = VioletRed3!20!black,
		colbacktitle = VioletRed2!30!white,
		colframe = VioletRed4,
		colback = VioletRed1!03,
		sharp corners = all,
		toprule = 0.000 mm,
		bottomrule = 0.500 mm,
		leftrule = 0.500 mm,
		rightrule = 0.000 mm,
	}
	{thr}
\NewTcbTheorem
	[
	number within = section
	]
	{proposition}
	{Proposition}
	{
		breakable,
		skin = enhanced,
		enhanced jigsaw,			
		attach boxed title to top left = { xshift = -5.000 mm, yshift = 0.000 mm },
		boxed title style = { boxrule = 0.000 pt, sharp corners = all },
		varwidth boxed title,
		fonttitle = \bfseries,
		coltitle = cyan5!30!black,
		colbacktitle = cyan7!50!white,
		colframe = cyan5,
		colback = cyan9!07,
		sharp corners = all,
		toprule = 0.000 mm,
		bottomrule = 0.500 mm,
		leftrule = 0.500 mm,
		rightrule = 0.000 mm,
	}
	{prp}

\newcounter{mathseed}
\pgfmathsetseed{\arabic{mathseed}}
\pgfdeclarelayer{background}
\pgfsetlayers{background, main}
\pgfdeclaredecoration{irregular fractal line}{init}
{
	\state{init}[width=\pgfdecoratedinputsegmentremainingdistance]
	{
		\pgfpathlineto{\pgfpoint{random*\pgfdecoratedinputsegmentremainingdistance}{(random*\pgfdecorationsegmentamplitude-0.02)*\pgfdecoratedinputsegmentremainingdistance}}
		\pgfpathlineto{\pgfpoint{\pgfdecoratedinputsegmentremainingdistance}{0pt}}
	}
}
\def\tornpaper#1{%
	\ifthenelse{\isodd{\value{mathseed}}}
	{%
		\tikz
		{
			\node[inner sep = 1em] (A) {#1};		
			\begin{pgfonlayer}{background}			
				\fill[paper]						
				\pgfextra{\pgfmathsetseed{\arabic{mathseed}}\addtocounter{mathseed}{1}}%
				{decorate[irregular cloudy border]{decorate{decorate{decorate{decorate[ragged border]{
										(A.north west) -- (A.north east)
				}}}}}}
				-- (A.south east)
				\pgfextra{\pgfmathsetseed{\arabic{mathseed}}}%
				{decorate[irregular spiky border]{decorate{decorate{decorate{decorate[ragged border]{
										-- (A.south west)
				}}}}}}
				-- (A.north west);
			\end{pgfonlayer}
		}
	}
	{%
		\tikz{
			\node[inner sep=1em] (A) {#1};  
			\begin{pgfonlayer}{background}  
				\fill[paper] 
				\pgfextra{\pgfmathsetseed{\arabic{mathseed}}\addtocounter{mathseed}{1}}%
				{decorate[irregular spiky border]{decorate{decorate{decorate{decorate[ragged border]{
										(A.north east) -- (A.north west)
				}}}}}}
				-- (A.south west)
				\pgfextra{\pgfmathsetseed{\arabic{mathseed}}}%
				{decorate[irregular cloudy border]{decorate{decorate{decorate{decorate[ragged border]{
										-- (A.south east)
				}}}}}}
				-- (A.north east);
		\end{pgfonlayer}}
	}
}
\title
	{
		Quantum Classification Outside the Promised Class
	}

\usepackage{orcidlink}
\newcommand{\orcidicon}[1]{\href{https://orcid.org/#1}{\includegraphics[height=\fontcharht\font`\B]{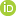}}}

\author
{
	Theodore Andronikos$^1$\orcidicon{0000-0002-3741-1271},
	Constantinos Bitsakos$^2$\orcidicon{0009-0003-3669-0453},
	Konstantinos Nikas$^2$\orcidicon{0000-0003-4424-9951},
	\\
	Georgios I. Goumas$^2$\orcidicon{0000-0001-7811-4831}
	and
	Nectarios Koziris$^2$\orcidicon{0000-0002-4890-8427}
	\\
	$^1$ \
	Department of Informatics, Ionian University, \\
	7 Tsirigoti Square, 49100 Corfu, Greece; \\
	andronikos@ionio.gr
	\\
	$^2$ \ 
	Computing Systems Laboratory, \\
	National Technical University of Athens, Greece; \\
	\{kbitsak, knikas, goumas, nkoziris\}@cslab.ece.ntua.gr
}
\begin{document}

\maketitle

\begin{abstract}
	This paper studies the important problem of quantum classification of Boolean functions from a entirely novel perspective. Typically, quantum classification algorithms allow us to classify functions with a probability of $1.0$, if we are promised that they meet specific unique properties. The primary objective of this study is to explore whether it is feasible to obtain any insights when the input function deviates from the promised class. For concreteness, we use a recently introduced quantum algorithm that is designed to classify with probability $1.0$ using just a single oracular query a large class of imbalanced Boolean functions. Fist, we establish a completely new concept characterizing ``nearness'' between Boolean function. Utilizing this concept, we show that, as long as the input function is close enough to the promised class, it is still possible to obtain useful information about its behavioral pattern from the classification algorithm. In this regard, the current study is among the first to provide evidence that shows how useful it is to apply quantum classification algorithms to functions outside the promised class in order to get a glimpse of important information.
\end{abstract}
\section{Introduction} \label{sec: Introduction}

In the realm of quantum computing, let us first recognize that there aren't any quantum computers that are strong enough to upend the status quo at this time. However, given the astounding developments in recent years, this is probably going to change sooner rather than later. Over the past few months, the majority of the key players in this field have made remarkable strides.
Following the success of the 127-qubit Eagle \cite{IBMEagle2021} and the 433-qubit Osprey \cite{IBMOsprey2022}, IBM has now introduced the 1,121-qubit Condor \cite{IBMCondor2023} and the newest and most powerful R2 Heron \cite{IBMHeron2024}.
Google has recently showcased the capability of quantum computers to surpass the performance of the most advanced supercomputers available today, as referenced in \cite{GoogleWillow2024,NatureGoogle2024}.
Microsoft has unveiled the groundbreaking Majorana 1 quantum chip, as noted in \cite{Aasen2025,Aghaee2025,MicrosoftMajorana2025} that promises to be a game-changer.
Regarding quantum annealers, D-Wave announced that they used a quantum processor to solve the first scientifically significant problem more quickly than they could have with a classical computer \cite{King2025, NatureD-Wave2025}. At the same time, other research teams have reported significant breakthroughs. Although this report is far from exhaustive, we believe that some of the most significant recent advancements are as follows.
The state-of-the-art Zuchongzhi 3.0 quantum processor, featuring 105 qubits, represents a significant technological achievement \cite{Gao2025,APSNewsZuchongzhi3.0-2025}.
Other significant developments include design concepts \cite{Cacciapuoti2024, Illiano2024} and hardware \cite{Photonic2024, NuQuantum2024}. Researchers recently demonstrated distributed quantum computing by successfully connecting two separate quantum processors using a photonic network interface, resulting in a single, fully integrated quantum computer \cite{Main2025, OxfordNewsEvents2025}. We believe that distributed quantum computing is currently becoming a reality.

This work examines the significant issue of quantum classification of Boolean functions from a completely new angle. It is generally possible to classify functions using quantum classification algorithms with a probability of $1.0$ if we are promised that they satisfy certain special characteristics. This study's main goal is to determine whether it is possible to gain any understanding when the input function deviates from the promised class. Staring from the ground-breaking Deutsch-Jozsa algorithm \cite{Deutsch1992}, numerous sophisticated works for quantum classification have appeared. The authors in \cite{Cleve1998} investigated a multidimensional variant of the Deutsch–Jozsa problem, which was subsequently elaborated upon in \cite{Chi2001} through the analysis of evenly distributed and balanced functions. Further advancements were made in \cite{Holmes2003}, where the Deutsch–Jozsa algorithm was adapted for balanced functions within finite Abelian subgroups. Another significant generalization was introduced in \cite{Ballhysa2004}. In \cite{Qiu2018}, researchers expanded the Deutsch–Jozsa problem and proposed an optimal algorithm. A more recent and innovative generalization of the Deutsch–Jozsa algorithm is presented in \cite{OssorioCastillo2023}. Additionally, practical applications of the Deutsch–Jozsa algorithm were explored in \cite{Qiu2020} and \cite{Zhengwei2021}. Notably, two significant contributions towards developing a distributed version of the Deutsch–Jozsa algorithm can be found in \cite{Tanasescu2019} and \cite{Li2025}. In a related context, the authors in \cite{Nagata2020} extended Deutsch’s algorithm to binary Boolean functions. It is also important to note that oracular algorithms designed for computing Boolean functions or for classification purposes are frequently discussed in the literature on Quantum Learning and Quantum Machine Learning, with notable studies including \cite{Bshouty1998}, \cite{Farhi1999}, \cite{Servedio2004}, \cite{Hunziker2009}, \cite{Yoo2014}, and \cite{Cross2015}. The vast majority of these works focus on the differentiation between constant and balanced functions, which are defined as functions where the number of elements in their domain that yield the value $0$ is equal to those that yield the value $1$. In contrast to this trend, a very recent study \cite{Andronikos2025a} focuses on imbalanced functions with an emphasis on classifying a large class of such functions exhibiting specific behavioral patterns with a novel quantum algorithm termed BFPQC for Boolean Function Pattern Quantum Classifier.

Our algorithm is presented as a game with the well-known Alice and Bob as characters. Games' entertaining elements are expected to help people comprehend the technical ideas more clearly. Since 1999 \cite{Meyer1999,Eisert1999}, quantum games have become increasingly popular because quantum strategies frequently perform better than classical ones \cite{Andronikos2018,Andronikos2021,Andronikos2022a}. This is best demonstrated by the well-known Prisoners' Dilemma \cite{Eisert1999}, which is a good example that can be applied to a number of other abstract quantum games \cite{Giannakis2015a,Koh2024}. Although quantum games seem like fun, they have been used to solve serious problems like like Quantum Key Distribution, Quantum Secret Sharing, Quantum Private Comparison (see \cite{Bennett1984,Ampatzis2021, Ampatzis2022, Ampatzis2023, Andronikos2023, Andronikos2023a, Andronikos2023b, Karananou2024, Andronikos2024, Andronikos2024a, Andronikos2024b, Andronikos2025}). Many classical systems can be transformed into quantum versions, including political frameworks, as demonstrated in recent studies \cite{Andronikos2022}. When it comes to games that take place in unusual settings, it is important to note that games that feature biological systems have attracted a lot of attention \cite{Theocharopoulou2019,Kastampolidou2020a,Kostadimas2021}. The fact that biosystems can produce biostrategies that might perform better than conventional strategies—even in the well-known Prisoners' Dilemma game—is especially fascinating \cite{Kastampolidou2020,Kastampolidou2021,Papalitsas2021,Kastampolidou2023,Adam2023}.

\textbf{Contribution}.
Generally, by utilizing a quantum classification algorithm, we can classify functions that meet specific properties with a probability of $1.0$. In the literature, it is common to encounter a phrase indicating that we have been assured the classifiable functions possess these special properties, leading to the term ``promise'' problem. In this study, we define the 'promised' class to include all functions that meet these criteria. Our analysis provides a precise definition of the promised class in Definition \ref{def: The Promised Class}. To eliminate any potential confusion, we emphasize that the classification algorithm is not anticipated to perform accurately, or more precisely, it is unlikely to yield the correct result when the input function does not belong to the promised class. In terms of implementation, these quantum algorithms utilize an oracle that discreetly executes the input function.

The main goal of this research is to determine whether it is possible to gain any understanding when the input function does not belong to the designated class. For this endeavor to produce significant outcomes, it needs to be clearly defined. As such, we operate based on the following assumptions. We fix the quantum classification algorithm BFPQC, which was first presented in \cite{Andronikos2025a}, and is capable of classifying all Boolean functions that belong to the promised class optimally and conclusively with a probability of $1.0$ using a single oracular query.
In this work we examine its behavior on functions that, although not belonging to the promised class, are still fairly close to its elements, as we will clarify in Definition \ref{def: Left & Right Clusters}. To achieve our goals, we present a brand-new idea that defines ``nearness'' between Boolean functions. By applying this idea, we demonstrate that the classification algorithm can still provide valuable insights into the input function's behavioral pattern as long as it is sufficiently close to the promised class. As far as we know, the current study is the first that offers proof of the value of applying quantum classification algorithms to functions outside of the promised class in order to gain a sneak peek at crucial information.

\subsection*{Organization} \label{subsec: Organization}

This article is structured in the following way. Section \ref{sec: Introduction} introduces the topic and includes references to relevant literature. Section \ref{sec: Background Concepts} offers a brief overview of key concepts, which serves as a basis for grasping the technical details. Sections \ref{sec: The Promised Class} and \ref{sec: Outside the Promised Class} rigorously define the promised class itself, and what we perceive as Boolean functions being outside the promised class, but still close enough to it. Section \ref{sec: What Did Alice Learn?} gives an in-depth analysis of the information that Alice can infer about the behavioral pattern of the unknown function. Finally, the paper wraps up with a summary and a discussion of the algorithm's nuances in Section \ref{sec: Discussion and Conclusions}.

\section{Background concepts} \label{sec: Background Concepts}

To begin, we will establish the notation and terminology that will be utilized throughout this article.

\begin{itemize}
	\item	
	$\mathbb{ B }$ is the binary set $\{ 0, 1 \}$.
	\item	
	$\mathbf{ b } = b_{ n - 1 } \dots b_{ 0 }$ is a sequence of $n$ bits that makes up a \emph{bit vector} $\mathbf{ b }$ of length $n$. \emph{zero} and \emph{one} are two special bit vectors, represented by $\mathbf{ 0 }$ and $\mathbf{ 1 }$, respectively, where all the bits are zero and one: $\mathbf{ 0 } = 0 \dots 0$ and $\mathbf{ 1 } = 1 \dots 1$.
	\item	
	For clarity, we write $\mathbf{ b }$ in boldface when referring to a bit vector $\mathbf{ b } \in \mathbb{ B }^{ n }$. For convenience, $\mathbf{ b }$ is frequently viewed as the binary representation of the integer $b$.
	\item	
	It is also possible to think of each bit vector $\mathbf{ b } \in \mathbb{ B }^{ n }$ as the binary correspondence to one of the $2^{ n }$ basis kets that constitute the computational basis of the $2^{ n }$-dimensional Hilbert space.
\end{itemize}

\begin{definition} {Boolean Function} { Boolean Function}
	A \emph{Boolean} function $f$ is a function from $\mathbb{ B }^{ n }$ to $\mathbb{ B }$, $n \geq 1$.
\end{definition}

Oracles represent a fundamental concept in quantum computing, serving as components in numerous quantum algorithms. An oracle functions as a black box that encapsulates a particular function or information within a quantum circuit, enabling quantum algorithms to address problems more effectively than classical algorithms in specific scenarios. It facilitates the evaluation of the function or the verification of a condition without disclosing the internal mechanisms of the function. In the realm of quantum algorithms, oracles are frequently employed to identify solutions to a problem or to convey insights regarding a function's behavior. For the purposes of this discussion, the following definition is adequate.

\begin{definition} {Oracle \& Unitary Transform} { Oracle & Unitary Transform}
	An \emph{oracle} is a black box implementing a Boolean function $f$. The idea here is that, being a black box function, we know nothing about its inner working; just that it works correctly. Thus, it can be used for the construction of a unitary transform $U_{ f }$ that captures the behavior of $f$.

	Henceforth, we shall assume that the corresponding unitary transform $U_{ f }$ implements the standard schema
	\begin{align}
		\label{eq: Generic Unitary Transform U_f}
		U_{ f }
		\colon
		\ket{ y }
		\
		\ket{ \mathbf{ x } }
		\rightarrow
		\ket{ y \oplus f( \mathbf{ x } ) }
		\
		\ket{ \mathbf{ x } }
		\ .
	\end{align}
\end{definition}

This particular oracle is occasionally designated as a Deutsch-Jozsa oracle. It is pertinent to mention that there exist other types of oracles, such as the Grover oracle, primarily employed to identify solutions to a problem. The present study assumes that every oracle and unitary transform adheres to Equation \eqref{eq: Generic Unitary Transform U_f} and is utilized to infer a function based on its behavior. The primary measure of complexity in oracular algorithms is the query complexity, defined as the number of queries directed towards the oracle by the algorithm.

For completeness, we recall the states $\ket{ + }$ and $\ket{ - }$, which are defined as

\begin{tcolorbox}
	[
		enhanced,
		breakable,
		center title,
		fonttitle = \bfseries,
		grow to left by = 0.000 cm,
		grow to right by = 0.000 cm,
		colback = white,			
		enhanced jigsaw,			
		sharp corners,
		toprule = 0.001 pt,
		bottomrule = 0.001 pt,
		leftrule = 0.001 pt,
		rightrule = 0.001 pt,
	]
	\begin{minipage} [ b ] { 0.450 \textwidth }
		\begin{align}
			\label{eq: Ket +}
			\ket{ + }
			=
			H
			\ket{ 0 }
			=
			\frac
			{ \ket{ 0 } + \ket{ 1 } }
			{ \sqrt{ 2 } }
		\end{align}
	\end{minipage}
	\begin{minipage} [ b ] { 0.450 \textwidth }
		\begin{align}
			\label{eq: Ket -}
			\ket{ - }
			=
			H
			\ket{ 1 }
			=
			\frac
			{ \ket{ 0 } - \ket{ 1 } }
			{ \sqrt{ 2 } }
		\end{align}
	\end{minipage}
\end{tcolorbox}

In the rest of this work, we shall often refer to following signature state $\ket{ \varphi }$

\begin{align}
	\label{eq: The Perfect SuperPosition State}
	\ket{ \varphi }
	&=
	2^{ - \frac { n } { 2 } }
	\sum_{ \mathbf{ x } \in \mathbb{ B }^{ n } }
	\ket{ \mathbf{ x } }
	\ ,
\end{align}

as the \emph{perfect superposition state}. This state, which is the staring point in most quantum algorithms, is typically obtained by acting with the $n$-fold Hadamard transform on the initial state $\ket{ 0 }^{ \otimes n }$.

To obtain any useful information from the schema \eqref{eq: Generic Unitary Transform U_f}, we set $\ket{ y }$ equal to $\ket{ - }$, in which case \eqref{eq: Generic Unitary Transform U_f} takes the following familiar form:

\begin{align}
	\label{eq: Unitary Transform U_f}
	U_{ f }
	\colon
	\ket{ - }
	\
	\ket{ \mathbf{ x } }
	\rightarrow
	( - 1 )^{ f ( \mathbf{ x } ) }
	\
	\ket{ - }
	\
	\ket{ \mathbf{ x } }
	\ .
\end{align}

Figures \ref{fig: The Quantum Circuit for the Generic Unitary Transform U_f} and \ref{fig: The Quantum Circuit for the Unitary Transform U_f} give a visual outline of the unitary transforms $U_{ f }$ that implement schemata \eqref{eq: Generic Unitary Transform U_f} and \eqref{eq: Unitary Transform U_f}, respectively.

\begin{tcolorbox}
	[
		enhanced,
		breakable,
		center title,
		fonttitle = \bfseries,
		grow to left by = 0.000 cm,
		grow to right by = 0.000 cm,
		colback = MagentaLighter!03,
		enhanced jigsaw,			
		sharp corners,
		toprule = 0.001 pt,
		bottomrule = 0.001 pt,
		leftrule = 0.001 pt,
		rightrule = 0.001 pt,
	]
	\begin{minipage} [ b ] { 0.450 \textwidth }
		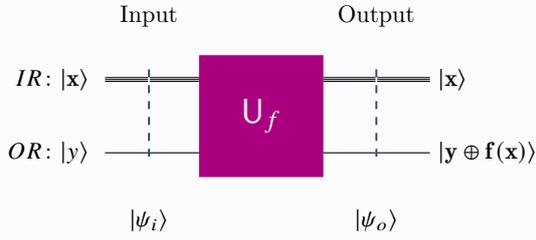
\begin{figure}[H]
			\centering
			\begin{tikzpicture} [ scale = 0.900 ] 
				\begin{yquant}
					qubits { $IR \colon \ket{ \mathbf{ x } }$ } IR;
					qubit { $OR \colon \ket{ y }$ } OR;
					[ name = Input, WordBlueDarker, line width = 0.250 mm, label = { [ label distance = 0.600 cm ] north: Input } ]
					barrier ( - ) ;
					hspace { 0.100 cm } IR;
					[ draw = RedPurple, fill = RedPurple, x radius = 0.900 cm, y radius = 0.700 cm ] box { \color{white} \Large \sf{U}$_{ f }$} (-);
					[ name = Output, WordBlueDarker, line width = 0.250 mm, label = { [ label distance = 0.600 cm ] north: Output } ]
					barrier ( - ) ;
					output { $\ket{ \mathbf{ x } }$ } IR;
					output { $\ket{ \mathbf{ y \oplus f ( \mathbf{ x } ) } }$ } OR;
					\node [ below = 1.250 cm ] at (Input) { $\ket{ \psi_{ i } }$ };
					\node [ below = 1.250 cm ] at (Output) { $\ket{ \psi_{ o } }$ };
				\end{yquant}
				\node [ anchor = center, below = 1.250 cm of Input ] (PhantomNode) { };
			\end{tikzpicture}
			\caption{This figure shows the unitary transform $U_{ f }$, which is based on the oracle for the function $f$ and implements the standard schema \eqref{eq: Generic Unitary Transform U_f}.}
			\label{fig: The Quantum Circuit for the Generic Unitary Transform U_f}
		\end{figure}
	\end{minipage}
	\hspace{ 0.750 cm }
	\begin{minipage} [ b ] { 0.450 \textwidth }
		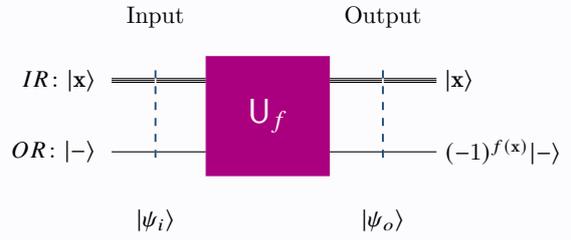
\begin{figure}[H]
			\centering
			\begin{tikzpicture} [ scale = 0.900 ] 
				\begin{yquant}
					qubits { $IR \colon \ket{ \mathbf{ x } }$ } IR;
					qubit { $OR \colon \ket{ - }$ } OR;
					[ name = Input, WordBlueDarker, line width = 0.250 mm, label = { [ label distance = 0.600 cm ] north: Input } ]
					barrier ( - ) ;
					hspace { 0.100 cm } IR;
					[ draw = RedPurple, fill = RedPurple, x radius = 0.900 cm, y radius = 0.700 cm ] box { \color{white} \Large \sf{U}$_{ f }$} (-);
					[ name = Output, WordBlueDarker, line width = 0.250 mm, label = { [ label distance = 0.600 cm ] north: Output } ]
					barrier ( - ) ;
					output { $\ket{ \mathbf{ x } }$ } IR;
					output { $( - 1 )^{ f ( \mathbf{ x } ) } \ket{ - }$ } OR;
					\node [ below = 1.250 cm ] at (Input) { $\ket{ \psi_{ i } }$ };
					\node [ below = 1.250 cm ] at (Output) { $\ket{ \psi_{ o } }$ };
				\end{yquant}
				\node [ anchor = center, below = 1.250 cm of Input ] (PhantomNode) { };
			\end{tikzpicture}
			\caption{This figure shows the unitary transform $U_{ f }$, again based on the oracle for the function $f$, but now implementing the schema \eqref{eq: Unitary Transform U_f}.}
			\label{fig: The Quantum Circuit for the Unitary Transform U_f}
		\end{figure}
	\end{minipage}
\end{tcolorbox}

In this work, the following conventions are applied to all quantum circuits, including those shown in Figures \ref{fig: The Quantum Circuit for the Generic Unitary Transform U_f} and \ref{fig: The Quantum Circuit for the Unitary Transform U_f}.

\begin{itemize}
	\item	
	The arrangement of the qubits follows the Qiskit convention \cite{Qiskit2025}, specifically the little-endian qubit indexing convention, which positions the least significant qubit at the top of the diagram and the most significant qubit at the bottom.
	\item	
	$IR$ is the quantum input register that contains a specific number of qubits.
	\item	
	$OR$ is the single-qubit output register that is initialized to an arbitrary state $\ket{ y }$ in Figure \ref{fig: The Quantum Circuit for the Generic Unitary Transform U_f} and to state $\ket{ - }$ in Figure \ref{fig: The Quantum Circuit for the Unitary Transform U_f}.
	\item	
	$U_{ f }$ is the unitary transform. Its precise mathematical expression depends on $f$ and is hidden. However, it is assumed that is satisfies relation \eqref{eq: Generic Unitary Transform U_f} in Figure \ref{fig: The Quantum Circuit for the Generic Unitary Transform U_f} and relation \eqref{eq: Unitary Transform U_f} in Figure \ref{fig: The Quantum Circuit for the Unitary Transform U_f}.
\end{itemize}

We point out that the term ``promise'' is frequently used in the literature to describe a specific property of the Boolean function $f$. This means that we are assured, or if you would prefer we are certain with probability $1.0$, that $f$ satisfies the property in question. One well-known example of this is the Deutsch–Jozsa algorithm, which promises that $f$ is either \emph{balanced} or \emph{constant}.

Extending the operation of addition modulo $2$ to bit vectors is a natural and fruitful generalization.

\begin{definition} {Bitwise Addition Modulo $2$} { Bitwise Addition Modulo $2$}
	Given two bit vectors $\mathbf{ x }, \mathbf{ y } \in \mathbb{ B }^{ n }$, with $\mathbf{ x } = x_{ n - 1 } \dots x_{ 0 }$ and $\mathbf{ y } = y_{ n - 1 } \dots y_{ 0 }$, we define their \emph{bitwise sum modulo} $2$, denoted by $\mathbf{ x } \oplus \mathbf{ y }$, as
	\begin{align}
		\label{eq: Bitwise Addition Modulo $2$}
		\mathbf{ x }
		\oplus
		\mathbf{ y }
		\coloneq
		( x_{ n - 1 } \oplus y_{ n - 1 } )
		\dots
		( x_{ 0 } \oplus y_{ 0 } )
		\ .
	\end{align}
\end{definition}

Following the standard approach, we use the same symbol $\oplus$ to denote the operation of addition modulo $2$ two between bits, and the bitwise sum modulo $2$ between two bit vectors because the context always makes clear the intended operation.

\section{The promised class} \label{sec: The Promised Class}

Typically, by employing a quantum classification algorithm, we are able to classify with probability $1.0$ functions that satisfy certain special properties. Most often in the literature one encounters some phrase which states that we have been given the promise that the classifiable functions fulfill these special properties. Hence, the term ``promise'' problem. In this work, we refer to the ``promised'' class to encompass all these functions that satisfy these properties. In our investigation, the promised class is rigorously defined in Definition \ref{def: The Promised Class}. To dispel any possible ambiguity, we clarify that the classification algorithm is not expected to work correctly, or to be more accurate, is improbable to give the right answer, whenever the input function fails outside the promised class. Implementation-wise these quantum algorithms make use of an oracle that stealthily implements the input function. The main premise of this work is to investigate if it possible to gain any information when the input function falls outside the promised class, as rigorously defined in Definition \ref{def: The Promised Class}. To bear any useful result, such an undertaking must be concrete. Therefore, we shall work under the following hypotheses.

\begin{enumerate}
	[ left = 0.600 cm, labelsep = 0.500 cm, start = 1 ]
	\renewcommand \labelenumi { (\textbf{H}$_{ \theenumi }$) }
	\item	We study the behavior of the quantum classification algorithm BFPQC, introduced in \cite{Andronikos2025a}. The acronym stands for Boolean Function Pattern Quantum Classifier.
	\item	BFPQC is designed to classify all Boolean functions belonging to the promised class with probability $1.0$ using just a single oracular query.
	\item	In this paper we study functions that despite being outside the promised class, they are still quite ``close'' to its elements, in the sense we shall make precise in Definition \ref{def: Left & Right Clusters}.
\end{enumerate}

Under the above hypotheses, the question we shall answer here is: can we get any useful information? And if so under what circumstances?

As previously mentioned, we conceptualize this method as the advancement of a game played between our talented stars, Alice and Bob. This game is called the Classification Game and is governed by the following rules.

\begin{enumerate}
	[ left = 0.600 cm, labelsep = 0.500 cm, start = 1 ]
	\renewcommand \labelenumi { (\textbf{G}$_{ \theenumi }$) }
	\item	Bob is free to choose any Boolean function, belonging either to the promised class, or to a class near to the promised class as ordained by Definition \ref{def: Left & Right Clusters}. Bob must announce to Alice the class to which the unknown function belongs.
	\item	Bob wins the game if Alice fails to identify the chosen function if it belongs to the promised class, or if Alice fails to obtain any useful information if it is outside the promised class. Otherwise, Alice is the winner.
	\item	In terms of implementing the game as a quantum circuit, Bob chooses the hidden oracle, while Alice furnishes the classifier, and is allowed to make a single query to the oracle.
\end{enumerate}

\begin{definition} {Pattern Bit Vector} { Pattern Vector}
	Let $f$ be a Boolean function from $\mathbb{ B }^{ n }$ to $\mathbb{ B }$, $n \geq 1$. We abstract the operation of $f$ into a unique bit vector that we call \emph{pattern bit vector}.The technicalities are given below.
	\begin{itemize}
		\item	
		The pattern bit vector $\mathbf{ p }_{ f }$ $=$ $p_{ 2^{ n } - 1 }$ $\dots$ $p_{ 1 }$ $p_{ 0 }$ corresponding to $f$ is the element of $\mathbb{ B }^{ 2^{ n } }$ with the property that $p_{ i } = f ( \mathbf{ i } )$, where $\mathbf{ i }$ is the binary bit vector representing integer $i$, $0 \leq i \leq 2^{ n } - 1$. The pattern bit vector $\mathbf{ p }_{ f }$ codifies the binary values of $f ( \mathbf{ i } )$ as $\mathbf{ i }$ ranges over $\mathbb{ B }^{ n }$.

				\begin{tblr}
					{
						colspec =
						{
							Q [ r, m, 4.000 cm ]
							Q [ c, m, 1.400 cm ]
							Q [ c, m, 0.350 cm ]
							Q [ c, m, 1.400 cm ]
							Q [ c, m, 0.000 cm ]
							Q [ c, m, 1.400 cm ]
						},
						rowspec =
						{
							Q
							Q
							Q
							Q
							Q
							Q
							Q
						}
					}
					\emph{Position:}
					&
					\SetCell { bg = VioletRed2, fg = black } $2^{ n } - 1$
					&
					\dots
					&
					\SetCell { bg = VioletRed2!50, fg = black } $1$
					&
					&
					\SetCell { bg = VioletRed2!25, fg = black } $0$
					\\
					&
					$\downarrow$
					&
					\dots
					&
					$\downarrow$
					&
					&
					$\downarrow$
					\\
					\emph{Position in binary:}
					&
					\SetCell { bg = SeaGreen3, fg = black } $\mathbf{ 1 \dots 1 1 }$
					&
					\dots
					&
					\SetCell { bg = SeaGreen3!50, fg = black } $\mathbf{ 0 \dots 0 1 }$
					&
					&
					\SetCell { bg = SeaGreen3!25, fg = black } $\mathbf{ 0 \dots 0 0 }$
					\\
					&
					$\downarrow$
					&
					\dots
					&
					$\downarrow$
					&
					&
					$\downarrow$
					\\
					\emph{Function value:}
					&
					\SetCell { bg = cyan7, fg = black } $f ( \mathbf{ 1 \dots 1 1 } )$
					&
					\dots
					&
					\SetCell { bg = cyan8, fg = black } $f ( \mathbf{ 0 \dots 0 1 } )$
					&
					&
					\SetCell { bg = cyan9, fg = black } $f ( \mathbf{ 0 \dots 0 0 } )$
					\\
					&
					$\downarrow$
					&
					\dots
					&
					$\downarrow$
					&
					&
					$\downarrow$
					\\
					\emph{Pattern bit:}
					&
					\SetCell { bg = LightSkyBlue3, fg = black } $p_{ 2^{ n } - 1 }$
					&
					\dots
					&
					\SetCell { bg = LightSkyBlue3!50, fg = black } $p_{ 1 }$
					&
					&
					\SetCell { bg = LightSkyBlue3!25, fg = black } $p_{ 0 }$
				\end{tblr}
		\item	
		The \emph{negation} of the pattern bit vector $\mathbf{ p }_{ f }$ $=$ $p_{ 2^{ n } - 1 }$ $\dots$ $p_{ 1 }$ $p_{ 0 }$, denoted by $\overline { \mathbf{ p }_{ f } }$, is the pattern bit vector $\overline { p_{ 2^{ n } - 1 } }$ $\dots$ $\overline { p_{ 1 } }$ $\overline { p_{ 0 } }$, which encodes the function $\overline { f }$.
	\end{itemize}
\end{definition}

\begin{definition} {Functions from Patterns} { Functions from Patterns}
	Conversely, to any bit vector $\mathbf{ p }$ $=$ $p_{ 2^{ n } - 1 }$ $\dots$ $p_{ 1 }$ $p_{ 0 }$ \ of length $2^{ n }$, we associate the Boolean function $f_{ \mathbf{ p } }$ with the property that $f ( \mathbf{ i } ) = p_{ i }$, where $\mathbf{ i }$ is the binary bit vector representing integer $i$, $0 \leq i \leq 2^{ n } - 1$.
\end{definition}

In the rest of this paper, when we want to refer to an arbitrary pattern bit vector without emphasizing the corresponding function $f$, we shall simply write $\mathbf{ p }$ instead of $\mathbf{ p }_{ f }$.

\begin{definition} {Pattern Ket} { Pattern Ket}
	To each pattern bit vector $\mathbf{ p }$ $=$ $p_{ 2^{ n } - 1 }$ $\dots$ $p_{ 1 }$ $p_{ 0 }$, we associate the $2^{ n }$-dimensional \emph{pattern ket} $\ket{ a_{ \mathbf{ p } } }$, with coefficients $a_{ i }$, $0 \leq i \leq 2^{ n } - 1$, given by the next formula.
	\begin{align}
		\label{eq: Pattern Bit Vector - Pattern Ket Coefficients Correlation}
		a_{ i }
		=
		\left\{
		\
		\begin{matrix*}[l]
			\phantom{ - } 2^{ - \frac { n } { 2 } } & \text{if } \ p_{ i } = 0 \ ,
			\\
			- 2^{ - \frac { n } { 2 } } & \text{if } \ p_{ i } = 1 \ .
		\end{matrix*}
		\right.
	\end{align}
\end{definition}

The visual interpretation of Definition \ref{def: Pattern Ket} is presented in Figure \ref{fig: Pattern Bit Vector - Pattern Ket Coefficients Correlation}.

\begin{tcolorbox}
	[
		enhanced,
		breakable,
		center title,
		fonttitle = \bfseries,
		grow to left by = 0.000 cm,
		grow to right by = 0.000 cm,
		colback = Purple1!03,
		enhanced jigsaw,							
		sharp corners,
		toprule = 0.001 pt,
		bottomrule = 0.001 pt,
		leftrule = 0.001 pt,
		rightrule = 0.001 pt,
	]
	\begin{figure}[H]
		\centering
		\begin{tikzpicture} 
			[
			scale = 1.000,
			arraynode/.style = { rectangle, minimum width = 1.200 cm, minimum height = 0.750 cm, font = \sffamily },
			vdotsnode/.style = { font = \sffamily\large, inner sep = 0.000 pt },
			labelnode/.style = { rectangle, minimum width = 1.200 cm, minimum height = 0.750 cm, fill = MediumPurple1!50, font = \sffamily\bfseries, inner sep = 4.000 pt }
			]
			\node [ labelnode ] (BitLabel) { Bit };
			\node [ arraynode, below = 0.500 cm of BitLabel ] (B0) { $\mathbf{ 0 }$ };
			\node [ arraynode, below = 0.250 cm of B0 ] (B1) { $\mathbf{ 1 }$ };
			\node [ vdotsnode, below = 0.250 cm of B1 ] (vdots0) { $\vdots$ };
			\node [ arraynode, below = 0.250 cm of vdots0 ] (B2) { $\mathbf{ 2^{ n } - 2 }$ };
			\node [ arraynode, below = 0.250 cm of B2 ] (B3) { $\mathbf{ 2^{ n } - 1 }$ };
			\node [ arraynode, fill = LightSkyBlue3!20, right = 0.750 cm of B0 ] (p_0) { $p_{ 0 }$ };
			\node [ labelnode, above = 0.500 cm of p_0 ] (PBVLabel) { $\mathbf{ p }$ };
			\node [ arraynode, fill = LightSkyBlue3!40, below = 0.250 cm of p_0] (p_1) { $p_{ 1 }$ };
			\node [ vdotsnode, below = 0.250 cm of p_1 ] (vdots1) { $\vdots$ };
			\node [ arraynode, fill = LightSkyBlue3!60, below = 0.250 cm of vdots1 ] (p_2N-2) { $p_{ 2^{ n } - 2 }$ };
			\node [ arraynode, fill = LightSkyBlue3!80, below = 0.250 cm of p_2N-2 ] (p_2N-1) { $p_{ 2^{ n } - 1 }$ };
			\node [ arraynode, fill = cyan9, right = 1.500 cm of p_0 ] (a_0) { $a_{ 0 }$ };
			\node [ labelnode, above = 0.500 cm of a_0 ] (PKLabel) { $\ket{ a_{ \mathbf{ p } } }$ };
			\node [ arraynode, fill = cyan8, below = 0.250 cm of a_0 ] (a_1) { $a_{ 1 }$ };
			\node [ vdotsnode, below = 0.250 cm of a_1 ] (vdots2) { $\vdots$ };
			\node [ arraynode, fill = cyan7!75, below = 0.250 cm of vdots2 ] (a_2N-2) { $a_{ 2^{ n } - 2 }$ };
			\node [ arraynode, fill = cyan6!75, below = 0.250 cm of a_2N-2 ] (a_2N-1) { $a_{ 2^{ n } - 1 }$ };
			\node [ arraynode, right = 0.750 cm of a_0 ] (C0) { $\mathbf{ 0 }$ };
			\node [ labelnode, above = 0.500 cm of C0] (CoefLabel) { Coefficient };
			\node [ arraynode, below = 0.250 cm of C0] (C1) { $\mathbf{ 1 }$ };
			\node [ vdotsnode, below = 0.250 cm of C1] (vdots3) { $\vdots$ };
			\node [ arraynode, below = 0.250 cm of vdots3] (C2) { $\mathbf{ 2^{ n } - 2 }$ };
			\node [ arraynode, below = 0.250 cm of C2] (C3) { $\mathbf{ 2^{ n } - 1 }$ };
			\draw [ Stealth-Stealth, line width = 1.750 pt, VioletRed3 ] (p_0.east) -- (a_0.west);
			\draw [ Stealth-Stealth, line width = 1.750 pt, VioletRed3 ] (p_1.east) -- (a_1.west);
			\draw [ Stealth-Stealth, line width = 1.750 pt, VioletRed3 ] (p_2N-2.east) -- (a_2N-2.west);
			\draw [ Stealth-Stealth, line width = 1.750 pt, VioletRed3 ] (p_2N-1.east) -- (a_2N-1.west);
		\end{tikzpicture}
		\caption{This figure visualizes the correlation among the bits of the pattern bit vector and the coefficients of the pattern ket.}
		\label{fig: Pattern Bit Vector - Pattern Ket Coefficients Correlation}
	\end{figure}
\end{tcolorbox}

The intuition behind the above Definition \ref{def: Pattern Ket} is quite straightforward. If $O_{ f }$ is an oracle realizing the Boolean function $f$, then, by recalling \eqref{eq: The Perfect SuperPosition State}, we surmise that $\ket{ a_{ \mathbf{ p } } }$ results from the action of $O_{ f }$ onto the perfect superposition state $\ket{ \varphi }$. For future reference we make a rigorous note of this fact in the following relation:

\begin{align}
	\label{eq: The Fundamental Property of Pattern Ket}
	O_{ f }
	\left(
	\ket{ - }
	\ket{ \varphi }
	\right)
	=
	\ket{ - }
	\ket{ a_{ \mathbf{ p } } }
	\ .
\end{align}

The previous Definitions \ref{def: Pattern Vector} and \ref{def: Pattern Ket} clearly illustrate a one-to-one relationship between Boolean functions and pattern bit vectors and between pattern bit vectors and pattern kets. One should consider a Boolean function, its corresponding pattern bit vector, and the corresponding pattern ket as different facets of the same entity. Consequently, understanding the behavior of a Boolean function allows for the creation of its pattern bit vector, or its pattern ket, while any one of the pattern bit vector or the pattern ket, holds all the essential information required to recreate the Boolean function. The next Figure \ref{fig: Equivalence Of Concepts I} is meant to convey this strong correlation among these three entities.

\begin{tcolorbox}
	[
		enhanced,
		breakable,
		center title,
		fonttitle = \bfseries,
		grow to left by = 0.000 cm,
		grow to right by = 0.000 cm,
		colback = SteelBlue1!03,
		enhanced jigsaw,			
		sharp corners,
		toprule = 0.001 pt,
		bottomrule = 0.001 pt,
		leftrule = 0.001 pt,
		rightrule = 0.001 pt,
	]
	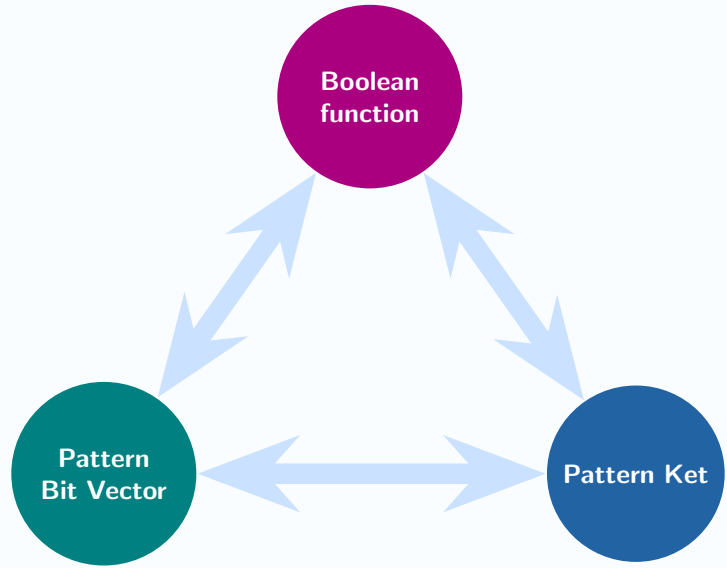
\begin{figure}[H]
		\centering
		\begin{tikzpicture} 
			[
			scale = 1.000,
			node distance = 1.000 cm,
			concept/.style =
			{
				circle,						
				thick,						
				inner sep = 1.000 pt,		
				text width = 2.250 cm,		
				align = center,				
				font = \sffamily,			
				text = white				
			}
			]
			\node [ concept, fill = RedPurple ] (A) at ( 0, 0 ) { \textbf{Boolean function} };
			\node [ concept, fill = GreenTeal ] (B) at ( -3.500, -5 ) { \textbf{Pattern Bit Vector} };
			\node [ concept, fill = azure4 ] (C) at ( 3.500, -5 ) { \textbf{Pattern Ket} };
			\draw [ Stealth-Stealth, line width = 7.750 pt, LightSteelBlue1 ] (A) -- (B);
			\draw [ Stealth-Stealth, line width = 7.750 pt, LightSteelBlue1 ] (A) -- (C);
			\draw [ Stealth-Stealth, line width = 7.750 pt, LightSteelBlue1 ] (B) -- (C);
		\end{tikzpicture}
		\caption{Boolean functions, pattern bit vectors and pattern kets are all equivalent.}
		\label{fig: Equivalence Of Concepts I}
	\end{figure}
\end{tcolorbox}

\begin{example} {Functions \& Patterns} { Functions & Patterns}

This example illustrates the previous definitions. Let us first consider the following binary Boolean function $f$:

\begin{align*}
	f
	(
	x_{ 1 },
	x_{ 0 }
	)
	\coloneq
	x_{ 1 }
	\wedge
	x_{ 0 }
	\ .
\end{align*}

The previous function takes the values listed in Table \ref{tbl: Truth Values & Pattern Bit Vector of 2D Boolean Function Example}. Accordingly, the corresponding pattern bit vector is $\mathbf{ p }_{ f } = 1000$.

\begin{tcolorbox}
	[
		enhanced,
		breakable,
		center title,
		fonttitle = \bfseries,
		grow to left by = 0.000 cm,
		grow to right by = 0.000 cm,
		colback = Purple1!03,
		enhanced jigsaw,							
		sharp corners,
		toprule = 0.001 pt,
		bottomrule = 0.001 pt,
		leftrule = 0.001 pt,
		rightrule = 0.001 pt,
	]
	\begin{table}[H]
		\caption{The truth values and the pattern bit vector of $f$.}
		\label{tbl: Truth Values & Pattern Bit Vector of 2D Boolean Function Example}
		\centering
		\SetTblrInner { rowsep = 1.000 mm }
		\begin{tblr}
			{
				colspec =
				{
					Q [ c, m, 1.000 cm ]
					| [ 1.000 pt, VioletRed3 ]
					| [ 1.000 pt, VioletRed3 ]
					Q [ c, m, 0.500 cm ]
					| [ 0.500 pt, VioletRed3 ]
					Q [ c, m, 0.500 cm ]
					| [ 0.500 pt, VioletRed3 ]
					Q [ c, m, 0.500 cm ]
					| [ 0.500 pt, VioletRed3 ]
					Q [ c, m, 0.500 cm ]
					| [ 1.000 pt, VioletRed3 ]
					| [ 1.000 pt, VioletRed3 ]
					Q [ c, m, 3.000 cm ]
				},
				rowspec =
				{
					|
					[ 3.500 pt, VioletRed3 ]
					|
					[ 0.750 pt, VioletRed3 ]
					|
					[ 0.250 pt, white ]
					Q
					|
					[ 0.500 pt, VioletRed3 ]
					Q
					|
					[ 3.500 pt, VioletRed3 ]
				}
			}
			&
			\SetCell { bg = VioletRed4, fg = white } $\mathbf{ 00 }$
			&
			\SetCell { bg = VioletRed4, fg = white } $\mathbf{ 01 }$
			&
			\SetCell { bg = VioletRed4, fg = white } $\mathbf{ 10 }$
			&
			\SetCell { bg = VioletRed4, fg = white } $\mathbf{ 11 }$
			&
			\SetCell { bg = VioletRed4, fg = white, font = \bfseries } {Pattern Bit Vector}
			\\
			$f$
			&
			$0$
			&
			$0$
			&
			$0$
			&
			$1$
			&
			$1 \ 0 \ 0 \ 0$
			\\
		\end{tblr}
	\end{table}
\end{tcolorbox}

If $O_{ f }$ is any oracle for $f$, its action on the perfect superposition state $\ket{ \varphi }$ results in the pattern ket $\ket{ a_{ \mathbf{ p } } }$, given below

\begin{align*}
	\ket{ a_{ \mathbf{ p } } }
	=
	\frac { 1 } { 2 } \ket{ 00 } +
	\frac { 1 } { 2 } \ket{ 01 } +
	\frac { 1 } { 2 } \ket{ 10 } -
	\frac { 1 } { 2 } \ket{ 11 }
	\ .
\end{align*}

\end{example}

\begin{definition} {Orthogonal Pattern Bit Vectors} { Orthogonal Pattern Bit Vectors}
	Given two pattern bit vectors $\mathbf{ p }$ and $\mathbf{ q }$ of length $2^{ n }$, we say that $\mathbf{ p }$ and $\mathbf{ q }$ are \emph{orthogonal} if \ $\mathbf{ p } \oplus \mathbf{ q }$ contains $2^{ n - 1 }$ $0$s and $2^{ n - 1 }$ $1$s.
\end{definition}

\begin{definition} {Pattern Bit Vectors Product} { Pattern Bit Vectors Product}
	Let $\mathbf{ p }$ $=$ $p_{ 2^{ n } - 1 }$ $\dots$ $p_{ 1 }$ $p_{ 0 }$ and $\mathbf{ q }$ be two pattern bit vectors of length $2^{ n }$ and $2^{ m }$, respectively, where, in general, $n \neq m$. We define their product, denoted by $\mathbf{ p } \odot \mathbf{ q }$, as
	\begin{align}
		\label{eq: Pattern Bit Vectors Product}
		&\mathbf{ p }
		\odot
		\mathbf{ q }
		=
		\mathbf{ s }_{ 2^{ n } - 1 } \dots \mathbf{ s }_{ 1 } \mathbf{ s }_{ 0 }
		\ ,
		\\
		&\text{where}
		\nonumber
		\\
		&\mathbf{ s }_{ i }
		=
		\begin{cases}
			\ \mathbf{ q }						& \text{if } p_{ i } = 0 
			\\[ 1.000 ex ]
			\ \overline{ \mathbf{ q } }			& \text{if } p_{ i } = 1 
		\end{cases}
		\ ,
		\quad
		0 \leq i \leq 2^{ n } - 1
		\ .
		\nonumber
	\end{align}
\end{definition}

\begin{definition} {Pattern Basis} { Pattern Basis}
	A \emph{pattern basis} of rank $n$, $n \geq 1$,
	is a list of $2^{ n }$ pairwise orthogonal pattern bit vectors $( \mathbf{ p }_{ 0 }, \mathbf{ p }_{ 1 }, \dots, \mathbf{ p }_{ 2^{ n } - 1 } )$, each of length $2^{ n }$.
\end{definition}

\begin{definition} {Pattern Bases Product} { Pattern Bases Product}
	Let $P_{ n }$ $=$ $( \mathbf{ p }_{ 0 }, \mathbf{ p }_{ 1 },$ \dots $, \mathbf{ p }_{ 2^{ n } - 1 } )$ and $P_{ m }^{ \prime }$ $=$ $( \mathbf{ q }_{ 0 },$ $\mathbf{ q }_{ 1 },$ $\dots $, $\mathbf{ q }_{ 2^{ m } - 1 } )$ be two pattern bases of rank $n$ and $m$, respectively, where, in general, $n \neq m$. Their product $P_{ n }$ $\odot$ $P_{ m }^{ \prime }$ is defined as
	\begin{align}
		\label{eq: Pattern Bases Product}
		P_{ n }
		\odot
		P_{ m }^{ \prime }
		\coloneq
		(
		\
		&\mathbf{ p }_{ 0 } \odot \mathbf{ q }_{ 0 },
		\mathbf{ p }_{ 0 } \odot \mathbf{ q }_{ 1 },
		\dots
		\mathbf{ p }_{ 0 } \odot \mathbf{ q }_{ 2^{ m } - 1 },
		\nonumber
		\\
		&\mathbf{ p }_{ 1 } \odot \mathbf{ q }_{ 0 },
		\mathbf{ p }_{ 1 } \odot \mathbf{ q }_{ 1 },
		\dots
		\mathbf{ p }_{ 1 } \odot \mathbf{ q }_{ 2^{ m } - 1 },
		\nonumber
		\\
		&\dots,
		\nonumber
		\\
		&\mathbf{ p }_{ 2^{ n } - 1 } \odot \mathbf{ q }_{ 0 },
		\mathbf{ p }_{ 2^{ n } - 1 } \odot \mathbf{ q }_{ 1 },
		\dots
		\mathbf{ p }_{ 2^{ n } - 1 } \odot \mathbf{ q }_{ 2^{ m } - 1 }
		\
		)
		\ ,
	\end{align}
	comprising a list of $2^{ n + m }$ pattern bit vectors, each of length $2^{ n + m }$.
\end{definition}

We employ the same symbol $\odot$ to denote the product operation between two pattern bit vectors, and the product operation between two pattern bases because the context always makes clear the intended operation.

Using Definitions \ref{def: Pattern Basis} and \ref{def: Pattern Bases Product}, we can immediately arrive at the next useful Theorem \ref{thr: Pattern Basis Construction}.

\begin{theorem} {Pattern Basis Construction} { Pattern Basis Construction}
	Let $P_{ n }$ and $P_{ m }^{ \prime }$ be two pattern bases of rank $n$ and $m$ (possibly $n \neq m$). Their product $P_{ n }$ $\odot$ $P_{ m }^{ \prime }$ is a pattern basis of rank $n + m$.
\end{theorem}

\begin{example} {Product Examples} { Product Examples}

We begin by giving two illustrative examples of the product operation between pattern bit vectors. To make clear all the details, we show the step-by-step product construction process in Tables \ref{tbl: The product $01 odot 1100$} and \ref{tbl: The product $0101 odot 0001$}.

\begin{align*}
	01
	\odot
	1100
	&=
	11000011
	\\
	0101
	\odot
	0001
	&=
	0001111000011110
\end{align*}

\begin{tcolorbox}
	[
		enhanced,
		breakable,
		center title,
		fonttitle = \bfseries,
		grow to left by = 0.000 cm,
		grow to right by = 0.000 cm,
		colback = white,							
		enhanced jigsaw,							
		sharp corners,
		toprule = 0.001 pt,
		bottomrule = 0.001 pt,
		leftrule = 0.001 pt,
		rightrule = 0.001 pt,
	]
	\begin{table}[H]
		\begin{minipage} [ t ] { 0.450 \textwidth }
			\caption{The product $01 \odot 1100$.}
			\label{tbl: The product $01 odot 1100$}
			\centering
			\SetTblrInner { rowsep = 1.000 mm }
			\begin{tblr}
				{
					colspec =
					{
						Q [ c, m, 1.750 cm ]
						Q [ c, m, 0.700 cm ]
						Q [ c, m, 0.700 cm ]
					},
					rowspec =
					{
						| [ 6.000 pt, white ]
						Q
						Q
						Q
						Q
						Q
						Q
						Q
						| [ 6.000 pt, white ]
					}
				}
				$\mathbf{ p } = 01$
				&
				\SetCell { bg = VioletRed1, fg = black } $0$
				&
				\SetCell { bg = VioletRed1!50, fg = black } $1$
				\\
				&
				$\downarrow$
				&
				$\downarrow$
				\\
				$\mathbf{ q } = 1100$
				&
				\SetCell { bg = SeaGreen2, fg = black } $1100$
				&
				\SetCell { bg = SeaGreen2!50, fg = black } $0011$
				\\
				&
				$\downarrow$
				&
				$\downarrow$
				\\
				\SetCell { bg = DarkSeaGreen1, fg = black } Product:
				&
				\SetCell [ c = 2 ] { c, bg = cyan8, fg = black } $1100 \quad 0011$
			\end{tblr}
		\end{minipage}
		\hspace{ 0.300 cm }
		\begin{minipage} [ t ] { 0.450 \textwidth }
			\caption{The product $0101 \odot 0001$.}
			\label{tbl: The product $0101 odot 0001$}
			\centering
			\SetTblrInner { rowsep = 1.000 mm }
			\begin{tblr}
				{
					colspec =
					{
						Q [ c, m, 1.750 cm ]
						Q [ c, m, 0.700 cm ]
						Q [ c, m, 0.700 cm ]
						Q [ c, m, 0.700 cm ]
						Q [ c, m, 0.700 cm ]
					},
					rowspec =
					{
						| [ 6.000 pt, white ]
						Q
						Q
						Q
						Q
						Q
						Q
						Q
						| [ 6.000 pt, white ]
					}
				}
				$\mathbf{ p } = 0101$
				&
				\SetCell { bg = Thistle3, fg = black } $0$
				&
				\SetCell { bg = Thistle3!80, fg = black } $1$
				&
				\SetCell { bg = Thistle3!60, fg = black } $0$
				&
				\SetCell { bg = Thistle3!40, fg = black } $1$
				\\
				&
				$\downarrow$
				&
				$\downarrow$
				&
				$\downarrow$
				&
				$\downarrow$
				\\
				$\mathbf{ q } = 0001$
				&
				\SetCell { bg = PaleTurquoise3, fg = black } $0001$
				&
				\SetCell { bg = PaleTurquoise3!80, fg = black } $1110$
				&
				\SetCell { bg = PaleTurquoise3!60, fg = black } $0001$
				&
				\SetCell { bg = PaleTurquoise3!40, fg = black } $1110$
				\\
				&
				$\downarrow$
				&
				$\downarrow$
				&
				$\downarrow$
				&
				$\downarrow$
				\\
				\SetCell { bg = Honeydew2, fg = black } Product:
				&
				\SetCell [ c = 4 ] { c, bg = azure7, fg = black } $0001 \quad 1110 \quad 0001 \quad 1110$
			\end{tblr}
		\end{minipage}
	\end{table}
\end{tcolorbox}

We now proceed to give an example of the product between two pattern bases. We focus on the pattern basis $I_{ 2 }$, which is the set consisting of the following four pairwise orthogonal pattern vectors:

\begin{align*}
	&I_{ 2 }
	\coloneq
	(
	\mathbf{ p }_{ 0 },
	\mathbf{ p }_{ 1 },
	\mathbf{ p }_{ 2 },
	\mathbf{ p }_{ 3 }
	)
	\ ,
	\\
	&\text{where}
	\\
	&\mathbf{ p }_{ 0 } = 0001,
	\mathbf{ p }_{ 1 } = 0010,
	\mathbf{ p }_{ 2 } = 0100,
	\mathbf{ p }_{ 3 } = 1000
	\ .
\end{align*}

The intuition behind our choice of the pattern basis $I_{ 2 }$ is that these pattern vectors correspond to imbalanced binary Boolean functions and can be used to generate a sequence of classes of imbalanced Boolean functions. We compute the product of $I_{ 2 }$ with itself

\begin{align*}
	I_{ 4 }
	\coloneq
	I_{ 2 }
	\odot
	I_{ 2 }
	\ .
\end{align*}

To show how Definition \ref{def: Pattern Bases Product} works, we list the elements of $I_{ 4 }$ in Table \ref{tbl: Pattern Basis $I_{ 4 }$}, indicating their derivation from $I_{ 2 }$.

\begin{tcolorbox}
	[
		enhanced,
		breakable,
		center title,
		fonttitle = \bfseries,
		grow to left by = 0.000 cm,
		grow to right by = 0.000 cm,
		colback = white,							
		enhanced jigsaw,							
		sharp corners,
		toprule = 0.001 pt,
		bottomrule = 0.001 pt,
		leftrule = 0.001 pt,
		rightrule = 0.001 pt,
	]
	\begin{table}[H]
		\caption{This table contains the pattern bit vectors of $I_{ 4 }$.}
		\label{tbl: Pattern Basis $I_{ 4 }$}
		\centering
		\SetTblrInner { rowsep = 1.200 mm }
		\begin{tblr}
			{
				colspec =
				{
					Q [ c, m, 2.000 cm ]
					| [ 0.750 pt, LightSkyBlue4 ]
					| [ 0.750 pt, LightSkyBlue4 ]
					Q [ c, m, 2.850 cm ]
					| [ 0.500 pt, LightSkyBlue4 ]
					Q [ c, m, 2.850 cm ]
					| [ 0.500 pt, LightSkyBlue4 ]
					Q [ c, m, 2.850 cm ]
					| [ 0.500 pt, LightSkyBlue4 ]
					Q [ c, m, 2.850 cm ]
				},
				rowspec =
				{
					| [ 3.500 pt, LightSkyBlue4 ]
					| [ 0.750 pt, LightSkyBlue4 ]
					| [ 0.250 pt, white ]
					Q
					|
					Q
					| [ 0.150 pt, LightSkyBlue4 ]
					Q
					| [ 0.150 pt, LightSkyBlue4 ]
					Q
					| [ 0.150 pt, LightSkyBlue4 ]
					Q
					| [ 3.500 pt, LightSkyBlue4 ]
				}
			}
			\SetCell { font = \bfseries \small } $\mathbf{ I }_{ 2 }$ Pattern Bit Vectors
			&
			\SetCell { bg = SkyBlue4, fg = white, font = \bfseries \small } $\mathbf{ p }_{ i } \odot \mathbf{ p }_{ 0 }, \ 0 \leq i \leq 3$
			&
			\SetCell { bg = SkyBlue4, fg = white, font = \bfseries \small } $\mathbf{ p }_{ i } \odot \mathbf{ p }_{ 1 }, \ 0 \leq i \leq 3$
			&
			\SetCell { bg = SkyBlue4, fg = white, font = \bfseries \small } $\mathbf{ p }_{ i } \odot \mathbf{ p }_{ 2 }, \ 0 \leq i \leq 3$
			&
			\SetCell { bg = SkyBlue4, fg = white, font = \bfseries \small } $\mathbf{ p }_{ i } \odot \mathbf{ p }_{ 3 }, \ 0 \leq i \leq 3$
			\\
			\SetCell { bg = SkyBlue4, fg = white, font = \bfseries \small } $\mathbf{ p }_{ 0 } = 0001$
			&
			{ \footnotesize $\mathbf{ r }_{ 0 } \colon 0001000100011110$ }
			&
			{ \footnotesize $\mathbf{ r }_{ 1 } \colon 0010001000101101$ }
			&
			{ \footnotesize $\mathbf{ r }_{ 2 } \colon 0100010001001011$ }
			&
			{ \footnotesize $\mathbf{ r }_{ 3 } \colon 1000100010000111$ }
			\\
			\SetCell { bg = SkyBlue4, fg = white, font = \bfseries \small } $\mathbf{ p }_{ 1 } = 0010$
			&
			{ \footnotesize $\mathbf{ r }_{ 4 } \colon 0001000111100001$ }
			&
			{ \footnotesize $\mathbf{ r }_{ 5 } \colon 0010001011010010$ }
			&
			{ \footnotesize $\mathbf{ r }_{ 6 } \colon 0100010010110100$ }
			&
			{ \footnotesize $\mathbf{ r }_{ 7 } \colon 1000100001111000$ }
			\\
			\SetCell { bg = SkyBlue4, fg = white, font = \bfseries \small } $\mathbf{ p }_{ 2 } = 0100$
			&
			{ \footnotesize $\mathbf{ r }_{ 8 } \colon 0001111000010001$ }
			&
			{ \footnotesize $\mathbf{ r }_{ 9 } \colon 0010110100100010$ }
			&
			{ \footnotesize $\mathbf{ r }_{ 10 } \colon 0100101101000100$ }
			&
			{ \footnotesize $\mathbf{ r }_{ 11 } \colon 1000011110001000$ }
			\\
			\SetCell { bg = SkyBlue4, fg = white, font = \bfseries \small } $\mathbf{ p }_{ 3 } = 1000$
			&
			{ \footnotesize $\mathbf{ r }_{ 12 } \colon 1110000100010001$ }
			&
			{ \footnotesize $\mathbf{ r }_{ 13 } \colon 1101001000100010$ }
			&
			{ \footnotesize $\mathbf{ r }_{ 14 } \colon 1011010001000100$ }
			&
			{ \footnotesize $\mathbf{ r }_{ 15 } \colon 0111100010001000$ }
			\\
		\end{tblr}
	\end{table}
\end{tcolorbox}

\end{example}

In view of the previous Definition \ref{def: Orthogonal Pattern Bit Vectors}, we may state the following immediate conclusion, in the form of Theorem \ref{thr: Orthogonal Pattern Kets}.

\begin{theorem} {Orthogonal Pattern Kets} { Orthogonal Pattern Kets}
	Let $\mathbf{ p }$ and $\mathbf{ q }$ be two orthogonal pattern bit vectors. Their corresponding pattern kets $\ket{ a_{ \mathbf{ p } } }$ and $\ket{ a_{ \mathbf{ q } } }$ are orthogonal:
	\begin{align}
		\label{eq: Orthogonal Pattern Kets}
		\braket
		{ a_{ \mathbf{ q } } }
		{ a_{ \mathbf{ p } } }
		=
		0
		\ .
	\end{align}
\end{theorem}

By construction the product operation between two pattern bit vectors, induces the following correlation among their corresponding pattern kets.

\begin{proposition} {Pattern Kets Tensor Product} { Pattern Kets Tensor Product}
	Let $\mathbf{ p }$, $\mathbf{ q }$, $\mathbf{ r }$ be pattern bit vectors, and $\ket{ a_{ \mathbf{ p } } }$, $\ket{ a_{ \mathbf{ q } } }$, and $\ket{ a_{ \mathbf{ r } } }$ their corresponding pattern kets. Then the following relation holds:
	\begin{align}
		\label{eq: Pattern Kets Tensor Product}
		\mathbf{ r }
		=
		\mathbf{ p }
		\odot
		\mathbf{ q }
		\Leftrightarrow
		\ket{ a_{ \mathbf{ r } } }
		=
		\ket{ a_{ \mathbf{ p } } }
		\otimes
		\ket{ a_{ \mathbf{ q } } }
		\ .
	\end{align}
\end{proposition}

\begin{definition} {Pattern Classifiers} { Pattern Classifiers}
	Let $P_{ n }$ $=$ $( \mathbf{ p }_{ 0 }, \mathbf{ p }_{ 1 }, \dots, \mathbf{ p }_{ 2^{ n } - 1 } )$ be a pattern basis of rank $n$, $n \geq 1$. To $P_{ n }$ corresponds the \emph{pattern classifier} $C_{ P_{ n } }$, defined as follows:
	\begin{align}
		\label{eq: Unitary Pattern Classifier}
		\NiceMatrixOptions
		{
			code-for-first-row = \color{GreenLighter2},
			cell-space-limits = 1.500 pt
		}
		C_{ P_{ n } }
		=
		\begin{bNiceMatrix}[ margin, first-row, hvlines, rules/width = 0.0025 mm ]	
			0^{ th } & 1^{ st } & \dots & ( 2^{ n } - 1 )^{ th }
			\\
			\ket{ a_{ \mathbf{ p_{ 0 } } } } & \ket{ a_{ \mathbf{ p_{ 1 } } } } & \dots & \ket{ a_{  \mathbf{ p }_{ 2^{ n } - 1 } } }
		\end{bNiceMatrix}
		\ ,
		\quad
		n \geq 1
		\ .
	\end{align}
\end{definition}

Taking into consideration Theorem \ref{thr: Orthogonal Pattern Kets} and Proposition \ref{prp: Pattern Kets Tensor Product}, we may surmise the next important fact.

\begin{theorem} {Pattern Classifiers Unitarity} { Pattern Classifiers Unitarity}
	If $C_{ P_{ n } }$ is the pattern classifier corresponding to the pattern basis $P_{ n }$, then $C_{ P_{ n } }$ is a unitary matrix.
\end{theorem}

In this work, we employ pattern classifier capable of identifying imbalanced Boolean functions. The primary characteristic of these functions is that the proportion of elements in their domain assigned the value $0$ is different from the proportion assigned the value $1$. The quantum algorithm BFPQC (Boolean Function Pattern Quantum Classifier), as introduced in \cite{Andronikos2025a}, classifies these functions with a probability of $1.0$ using a single oracular query. This article aims to explore the information that can be derived from the BFPQC algorithm when the input is a function that, while not part of this sequence, remains relatively similar. First, we formally define the aforementioned sequence of classes of imbalanced Boolean functions. we start from the pattern basis $I_{ 2 }$ that we have already encountered in Example \ref{xmp: Product Examples}. Let us recall that $I_{ 2 }$ consists of the following four pairwise orthogonal pattern bit vectors:

\begin{align}
	\label{eq: Pattern Basis I_2}
	&I_{ 2 }
	\coloneq
	(
	\mathbf{ p }_{ 0 },
	\mathbf{ p }_{ 1 },
	\mathbf{ p }_{ 2 },
	\mathbf{ p }_{ 3 }
	)
	\ ,
	\nonumber
	\\
	&\text{where}
	\nonumber
	\\
	&\mathbf{ p }_{ 0 } = 0001,
	\mathbf{ p }_{ 1 } = 0010,
	\mathbf{ p }_{ 2 } = 0100,
	\mathbf{ p }_{ 3 } = 1000
	\ .
\end{align}

\begin{definition} {A Sequence of Imbalanced Pattern Bases} { A Sequence Of Imbalanced Pattern Bases}
	We recursively define a sequence $( I_{ 2 n } )_{ n \geq 1 }$, of pattern bases exhibiting imbalance.
	\begin{enumerate}
		[ left = 0.500 cm, labelsep = 0.500 cm, start = 0 ]
		\renewcommand \labelenumi { $($\textbf{I}$_{ \theenumi }$$)$ }
		\item	If $n = 1$, the corresponding pattern basis is $I_{ 2 }$, as defined by \eqref{eq: Pattern Basis I_2}.
		\item	Given the pattern basis $I_{ 2 n }$ of rank $2 n$, we define the pattern basis $I_{ 2 ( n + 1 ) }$ of rank $2 ( n + 1 )$ as follows:
		\begin{align}
			\label{eq: Pattern Basis $I_{ 2 ( n + 1 ) }$}
			I_{ 2 ( n + 1 ) }
			\coloneq
			I_{ 2 }
			\odot
			I_{ 2 n }
			\ .
		\end{align}
		\item	This sequence of pattern bases $I_{ 2 }$, $I_{ 4 }$, \dots, $I_{ 2 n }$, \dots, is denoted by $\mathcal{ I }$.
	\end{enumerate}
\end{definition}

As we have pointed out, the sequence $\mathcal{ I }$ of pattern bases induces a corresponding sequence of classes of Boolean functions. This concept is formalized by Definition \ref{def: Functions From Imbalanced Bases}.

\begin{definition} {Functions from Imbalanced Bases} { Functions From Imbalanced Bases}
	Let $I_{ 2 n }$ $=$ $( \mathbf{ p }_{ 0 }, \mathbf{ p }_{ 1 },$ \dots $, \mathbf{ p }_{ 2^{ 2 n } - 1 } )$ be a pattern basis of rank $2 n$. To $I_{ 2 n }$ we associate the class $F_{ 2 n }$ $=$ $( f_{ 0 }, f_{ 1 },$ \dots $, f_{ 2^{ 2 n } - 1 } )$ of Boolean functions from $\mathbb{ B }^{ 2 n }$ to $\mathbb{ B }$ such that $\mathbf{ p }_{ i }$ is the pattern bit vector corresponding to $f_{ i }$, $0 \leq i \leq 2^{ 2 n } - 1$. This sequence of classes of Boolean functions $F_{ 2 }$, $F_{ 4 }$, \dots, $F_{ 2 n }$, \dots, is denoted by $\mathcal{ F }$.
\end{definition}

\begin{definition} {The Promised Class} { The Promised Class}
	The \emph{promised class} is the collection of Boolean functions belonging to the $\mathcal{ F }$ sequence.
\end{definition}


\begin{definition} {Pattern Classifiers for Imbalanced Functions} { Pattern Classifiers for Imbalanced Functions}
	We recursively define a sequence of unitary pattern classifiers corresponding to the sequence $\mathcal{ I }$ of imbalanced pattern bases as follows.
	\begin{enumerate}
		[ left = 0.750 cm, labelsep = 0.500 cm, start = 0 ]
		\renewcommand \labelenumi { $($\textbf{CI}$_{ \theenumi }$$)$ }
		\item	For $n = 1$, the unitary pattern classifier corresponding to $I_{ 2 }$ is denoted by $C_{ I_{ 2 } }$, and, in accordance to Definition \ref{def: Pattern Classifiers}, is given by
		\begin{align}
			\label{eq: Unitary Pattern Classifier $C_{ I_{ 2 } }$}
			\NiceMatrixOptions
			{
				code-for-first-row = \color{GreenLighter2},
				cell-space-limits = 1.500 pt
			}
			C_{ I_{ 2 } }
			=
			\begin{bNiceMatrix}[ margin, first-row, hvlines, rules/width = 0.0025 mm ]	
				0^{ th } & 1^{ st } & 2^{ nd } & 3^{ rd }
				\\
				\ket{ a_{ \mathbf{ p_{ 0 } } } } & \ket{ a_{ \mathbf{ p_{ 1 } } } } & \ket{ a_{ \mathbf{ p_{ 2 } } } } & \ket{ a_{  \mathbf{ p }_{ 3 } } }
			\end{bNiceMatrix}
			\ .
		\end{align}
		\item	Given the unitary pattern classifier $C_{ I_{ 2 n } }$ corresponding to the pattern basis $I_{ 2 n }$, we define $C_{ I_{ 2 ( n + 1 ) } }$, corresponding to $I_{ 2 ( n + 1 ) }$ as
		\begin{align}
			\label{eq: Unitary Pattern Classifier $C_{ I_{ 2 ( n + 1 ) } }$}
			C_{ I_{ 2 ( n + 1 ) } }
			\coloneq
			C_{ I_{ 2 } }
			\otimes
			C_{ I_{ 2 n } }
			=
			C_{ I_{ 2 } }^{ \otimes ( n + 1 ) }
			\qquad
			( n \geq 1 )
			\ .
		\end{align}
		\item	We use $\mathcal{ CI }$ for the sequence of unitary pattern classifiers $C_{ I_{ 2 } }$, $C_{ I_{ 4 } }$, \dots, $C_{ I_{ 2 n } }$, \dots \ .
	\end{enumerate}
\end{definition}

By combining Definition \ref{def: Pattern Classifiers for Imbalanced Functions} with Theorem \ref{thr: Orthogonal Pattern Kets} and Proposition \ref{prp: Pattern Kets Tensor Product}, we immediately arrive at the next Proposition \ref{prp: Pattern Classifier Action On Pattern Kets}.

\begin{proposition} {Pattern Classifier Action on Pattern Kets} { Pattern Classifier Action On Pattern Kets}
	Let $\ket{ a_{ \mathbf{ p_{ i } } } }$ be the $i^{ th }$ column of the unitary pattern classifier $C_{ I_{ 2 n } }$, $0 \leq i \leq 2^{ 2 n } - 1$. Then $C_{ I_{ 2 n } } \ket{ a_{ \mathbf{ p_{ i } } } } = \ket{ \mathbf{ i } }$, where $\mathbf{ i }$ is the binary representation of the index $i$, and $\ket{ \mathbf{ i } }$ is one of the basis kets of the computational basis.
\end{proposition}

The previous Definitions \ref{def: A Sequence Of Imbalanced Pattern Bases}, \ref{def: Functions From Imbalanced Bases}, and \ref{def: Pattern Classifiers for Imbalanced Functions} clearly illustrate a one-to-one correspondence among the elements of the pattern basis $I_{ 2 n }$, the functions of the function class $F_{ 2 n }$, and the column kets of the unitary pattern classifier $C_{ I_{ 2 n } }$ since they all contain precisely the same information.

\begin{tcolorbox}
	[
		enhanced,
		breakable,
		center title,
		fonttitle = \bfseries,
		grow to left by = 0.000 cm,
		grow to right by = 0.000 cm,
		colback = SteelBlue1!03,
		enhanced jigsaw,			
		sharp corners,
		toprule = 0.001 pt,
		bottomrule = 0.001 pt,
		leftrule = 0.001 pt,
		rightrule = 0.001 pt,
	]
	\begin{figure}[H]
		\centering
		\begin{tikzpicture} 
			[
			scale = 1.000,
			node distance = 3.000 cm,
			concept/.style =
			{
				rectangle,							
				rounded corners = 5.000 pt,			
				minimum width = 3.950 cm,
				minimum height = 1.750 cm,
				thick,								
				inner sep = 5.000 pt,				
				align = center,						
				font = \sffamily,					
				text = white						
			}
			]
			\node [ concept, fill = PaleVioletRed3 ] (A) at ( 0, 0 )
			{
				Pattern Bit Vector $\mathbf{ p }_{ i }$ \\
				in Pattern Basis \\
				$I_{ 2 n }$ $=$ $( \mathbf{ p }_{ 0 }, \mathbf{ p }_{ 1 }, \dots, \mathbf{ p }_{ 2^{ 2 n } - 1 })$
			};
			\node [ concept, fill = LightSteelBlue4 ] (B) at ( -4.250, -5.000 )
			{
				Boolean function $f_{ i }$ \\
				in function class \\
				$F_{ 2 n }$ $=$ $( f_{ 0 }, f_{ 1 },$ \dots $, f_{ 2^{ 2 n } - 1 } )$
			};
			\node [ concept, fill = CadetBlue4 ] (C) at ( 4.250, -5.000 )
			{
				Column ket $\ket{ a_{ \mathbf{ p_{ i } } } }$ \\
				in unitary pattern classifier \\
				$C_{ I_{ 2 n } }$
			};
			\draw [ Stealth-Stealth, line width = 7.750 pt, Azure3 ] (A) -- (B);
			\draw [ Stealth-Stealth, line width = 7.750 pt, Azure3 ] (A) -- (C);
			\draw [ Stealth-Stealth, line width = 7.750 pt, Azure3 ] (B) -- (C);
		\end{tikzpicture}
		\caption{One-to-one correspondence among the elements of the pattern basis $I_{ 2 n }$, the functions of the function class $F_{ 2 n }$, and the columns of the pattern classifier $C_{ I_{ 2 n ) } }$.}
		\label{fig: Equivalence Of Concepts II}
	\end{figure}
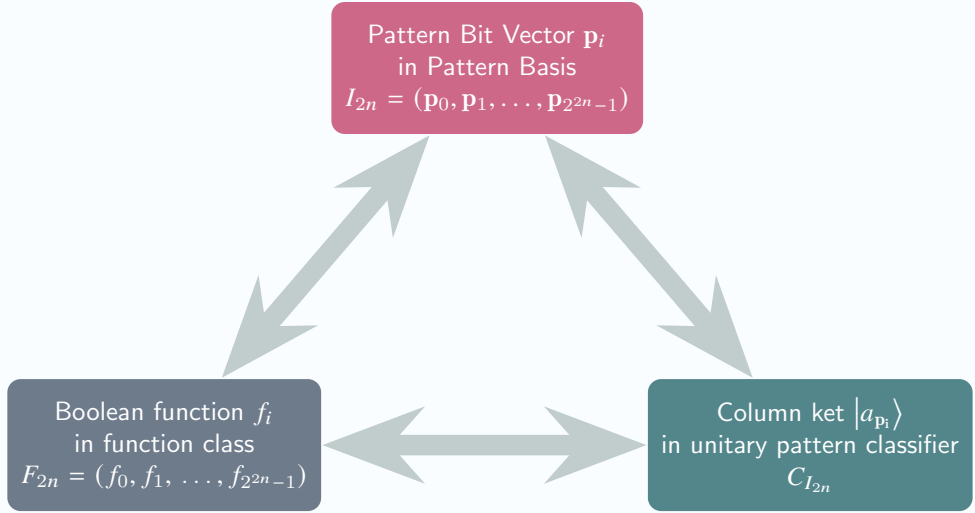
\end{tcolorbox}

\begin{example} {Unitary Classifiers Examples} { Unitary Classifiers Examples}

We start by explicitly writing the $2^{ 2 } \times 2^{ 2 }$ unitary classifier $C_{ I_{ 2 } }$ that corresponds to the pattern basis $I_{ 2 }$. Its precise construction follows the guidelines outlined in Definitions \ref{def: Pattern Ket} and \ref{def: Pattern Classifiers}. The end result is shown below:

\begin{align}
	\label{eq: Unitary Classifier $C_{ I_{ 2 } }$ Explicit Matrix Form}
	\NiceMatrixOptions
	{
		code-for-first-row = \color{GreenLighter2},
		cell-space-limits = 1.500 pt
	}
	C_{ I_{ 2 } }
	=
	\begin{bNiceMatrix}[ margin ] 
		- \frac { 1 } { 2 } & \phantom{-} \frac { 1 } { 2 } & \phantom{-} \frac { 1 } { 2 } & \phantom{-} \frac { 1 } { 2 } \\
		\phantom{-} \frac { 1 } { 2 } & - \frac { 1 } { 2 } & \phantom{-} \frac { 1 } { 2 } & \phantom{-} \frac { 1 } { 2 } \\
		\phantom{-} \frac { 1 } { 2 } & \phantom{-} \frac { 1 } { 2 } & - \frac { 1 } { 2 } & \phantom{-} \frac { 1 } { 2 } \\
		\phantom{-} \frac { 1 } { 2 } & \phantom{-} \frac { 1 } { 2 } & \phantom{-} \frac { 1 } { 2 } & - \frac { 1 } { 2 } \\
	\end{bNiceMatrix}
	=
	\begin{bNiceMatrix}[ margin, first-row, hvlines, rules/width = 0.0025 mm ]	
		0^{ th } & 1^{ st } & 2^{ nd } & 3^{ rd }
		\\
		\ket{ a_{ \mathbf{ p_{ 0 } } } } & \ket{ a_{ \mathbf{ p_{ 1 } } } } & \ket{ a_{ \mathbf{ p_{ 2 } } } } & \ket{ a_{  \mathbf{ p }_{ 3 } } }
	\end{bNiceMatrix}
	\ .
\end{align}

Incidentally, we note that $C_{ I_{ 2 } }$ can be easily constructed by modern quantum computers using ubiquitous standard gates such as the Hadamard, $Z$ and controlled-$Z$ gates:

\begin{align}
	\label{eq: Unitary Classifier $C_{ I_{ 2 } }$}
	C_{ I_{ 2 } }
	=
	( H \otimes H )
	\
	CZ
	\
	( Z \otimes Z )
	\
	( H \otimes H )
\end{align}

In the same way, we construct the matrix representation of the unitary classifier $C_{ I_{ 4 } }$, given below.

\begin{align}
	\label{eq: Unitary Classifier $C_{ I_{ 4 } }$ Explicit Matrix Form}
	C_{ I_{ 4 } }
	\overset { \eqref{eq: Unitary Pattern Classifier $C_{ I_{ 2 ( n + 1 ) } }$} } { = }
	C_{ I_{ 2 } }
	\otimes
	C_{ I_{ 2 } }
	\overset { \eqref{eq: Unitary Classifier $C_{ I_{ 2 } }$ Explicit Matrix Form} } { = }
	\NiceMatrixOptions{ cell-space-limits = 1.500 pt }
	\begin{bNiceMatrix}[ margin ] 
		- \frac { 1 } { 2 } & \phantom{-} \frac { 1 } { 2 } & \phantom{-} \frac { 1 } { 2 } & \phantom{-} \frac { 1 } { 2 } \\
		\phantom{-} \frac { 1 } { 2 } & - \frac { 1 } { 2 } & \phantom{-} \frac { 1 } { 2 } & \phantom{-} \frac { 1 } { 2 } \\
		\phantom{-} \frac { 1 } { 2 } & \phantom{-} \frac { 1 } { 2 } & - \frac { 1 } { 2 } & \phantom{-} \frac { 1 } { 2 } \\
		\phantom{-} \frac { 1 } { 2 } & \phantom{-} \frac { 1 } { 2 } & \phantom{-} \frac { 1 } { 2 } & - \frac { 1 } { 2 } \\
	\end{bNiceMatrix}
	\otimes
	\begin{bNiceMatrix}[ margin ] 
		- \frac { 1 } { 2 } & \phantom{-} \frac { 1 } { 2 } & \phantom{-} \frac { 1 } { 2 } & \phantom{-} \frac { 1 } { 2 } \\
		\phantom{-} \frac { 1 } { 2 } & - \frac { 1 } { 2 } & \phantom{-} \frac { 1 } { 2 } & \phantom{-} \frac { 1 } { 2 } \\
		\phantom{-} \frac { 1 } { 2 } & \phantom{-} \frac { 1 } { 2 } & - \frac { 1 } { 2 } & \phantom{-} \frac { 1 } { 2 } \\
		\phantom{-} \frac { 1 } { 2 } & \phantom{-} \frac { 1 } { 2 } & \phantom{-} \frac { 1 } { 2 } & - \frac { 1 } { 2 } \\
	\end{bNiceMatrix}
	\nonumber
	\\
	=
	{\footnotesize
		\begin{bNiceMatrix}[ margin, cell-space-limits = 1.000 pt ] 
			\noalign{\vskip 3pt}
			\phantom{-} \frac { 1 } { 4 } & - \frac { 1 } { 4 } & - \frac { 1 } { 4 } & - \frac { 1 } { 4 } &
			- \frac { 1 } { 4 } & \phantom{-} \frac { 1 } { 4 } & \phantom{-} \frac { 1 } { 4 } & \phantom{-} \frac { 1 } { 4 } &
			- \frac { 1 } { 4 } & \phantom{-} \frac { 1 } { 4 } & \phantom{-} \frac { 1 } { 4 } & \phantom{-} \frac { 1 } { 4 } &
			- \frac { 1 } { 4 } & \phantom{-} \frac { 1 } { 4 } & \phantom{-} \frac { 1 } { 4 } & \phantom{-} \frac { 1 } { 4 }
			\\
			- \frac { 1 } { 4 } & \phantom{-} \frac { 1 } { 4 } & - \frac { 1 } { 4 } & - \frac { 1 } { 4 } &
			\phantom{-} \frac { 1 } { 4 } & - \frac { 1 } { 4 } & \phantom{-} \frac { 1 } { 4 } & \phantom{-} \frac { 1 } { 4 } &
			\phantom{-} \frac { 1 } { 4 } & - \frac { 1 } { 4 } & \phantom{-} \frac { 1 } { 4 } & \phantom{-} \frac { 1 } { 4 } &
			\phantom{-} \frac { 1 } { 4 } & - \frac { 1 } { 4 } & \phantom{-} \frac { 1 } { 4 } & \phantom{-} \frac { 1 } { 4 }
			\\
			- \frac { 1 } { 4 } & - \frac { 1 } { 4 } & \phantom{-} \frac { 1 } { 4 } & - \frac { 1 } { 4 } &
			\phantom{-} \frac { 1 } { 4 } & \phantom{-} \frac { 1 } { 4 } & - \frac { 1 } { 4 } & \phantom{-} \frac { 1 } { 4 } &
			\phantom{-} \frac { 1 } { 4 } & \phantom{-} \frac { 1 } { 4 } & - \frac { 1 } { 4 } & \phantom{-} \frac { 1 } { 4 } &
			\phantom{-} \frac { 1 } { 4 } & \phantom{-} \frac { 1 } { 4 } & - \frac { 1 } { 4 } & \phantom{-} \frac { 1 } { 4 }
			\\
			- \frac { 1 } { 4 } & - \frac { 1 } { 4 } & - \frac { 1 } { 4 } & \phantom{-} \frac { 1 } { 4 } &
			\phantom{-} \frac { 1 } { 4 } & \phantom{-} \frac { 1 } { 4 } & \phantom{-} \frac { 1 } { 4 } & - \frac { 1 } { 4 } &
			\phantom{-} \frac { 1 } { 4 } & \phantom{-} \frac { 1 } { 4 } & \phantom{-} \frac { 1 } { 4 } & - \frac { 1 } { 4 } &
			\phantom{-} \frac { 1 } { 4 } & \phantom{-} \frac { 1 } { 4 } & \phantom{-} \frac { 1 } { 4 } & - \frac { 1 } { 4 }
			\\
			- \frac { 1 } { 4 } & \phantom{-} \frac { 1 } { 4 } & \phantom{-} \frac { 1 } { 4 } & \phantom{-} \frac { 1 } { 4 } &
			\phantom{-} \frac { 1 } { 4 } & - \frac { 1 } { 4 } & - \frac { 1 } { 4 } & - \frac { 1 } { 4 } &
			- \frac { 1 } { 4 } & \phantom{-} \frac { 1 } { 4 } & \phantom{-} \frac { 1 } { 4 } & \phantom{-} \frac { 1 } { 4 } &
			- \frac { 1 } { 4 } & \phantom{-} \frac { 1 } { 4 } & \phantom{-} \frac { 1 } { 4 } & \phantom{-} \frac { 1 } { 4 }
			\\
			\phantom{-} \frac { 1 } { 4 } & - \frac { 1 } { 4 } & \phantom{-} \frac { 1 } { 4 } & \phantom{-} \frac { 1 } { 4 } &
			- \frac { 1 } { 4 } & \phantom{-} \frac { 1 } { 4 } & - \frac { 1 } { 4 } & - \frac { 1 } { 4 } &
			\phantom{-} \frac { 1 } { 4 } & - \frac { 1 } { 4 } & \phantom{-} \frac { 1 } { 4 } & \phantom{-} \frac { 1 } { 4 } &
			\phantom{-} \frac { 1 } { 4 } & - \frac { 1 } { 4 } & \phantom{-} \frac { 1 } { 4 } & \phantom{-} \frac { 1 } { 4 }
			\\
			\phantom{-} \frac { 1 } { 4 } & \phantom{-} \frac { 1 } { 4 } & - \frac { 1 } { 4 } & \phantom{-} \frac { 1 } { 4 } &
			- \frac { 1 } { 4 } & - \frac { 1 } { 4 } & \phantom{-} \frac { 1 } { 4 } & - \frac { 1 } { 4 } &
			\phantom{-} \frac { 1 } { 4 } & \phantom{-} \frac { 1 } { 4 } & - \frac { 1 } { 4 } & \phantom{-} \frac { 1 } { 4 } &
			\phantom{-} \frac { 1 } { 4 } & \phantom{-} \frac { 1 } { 4 } & - \frac { 1 } { 4 } & \phantom{-} \frac { 1 } { 4 }
			\\
			\phantom{-} \frac { 1 } { 4 } & \phantom{-} \frac { 1 } { 4 } & \phantom{-} \frac { 1 } { 4 } & - \frac { 1 } { 4 } &
			- \frac { 1 } { 4 } & - \frac { 1 } { 4 } & - \frac { 1 } { 4 } & \phantom{-} \frac { 1 } { 4 } &
			\phantom{-} \frac { 1 } { 4 } & \phantom{-} \frac { 1 } { 4 } & \phantom{-} \frac { 1 } { 4 } & - \frac { 1 } { 4 } &
			\phantom{-} \frac { 1 } { 4 } & \phantom{-} \frac { 1 } { 4 } & \phantom{-} \frac { 1 } { 4 } & - \frac { 1 } { 4 }
			\\
			- \frac { 1 } { 4 } & \phantom{-} \frac { 1 } { 4 } & \phantom{-} \frac { 1 } { 4 } & \phantom{-} \frac { 1 } { 4 } &
			- \frac { 1 } { 4 } & \phantom{-} \frac { 1 } { 4 } & \phantom{-} \frac { 1 } { 4 } & \phantom{-} \frac { 1 } { 4 } &
			\phantom{-} \frac { 1 } { 4 } & - \frac { 1 } { 4 } & - \frac { 1 } { 4 } & - \frac { 1 } { 4 } &
			- \frac { 1 } { 4 } & \phantom{-} \frac { 1 } { 4 } & \phantom{-} \frac { 1 } { 4 } & \phantom{-} \frac { 1 } { 4 }
			\\
			\phantom{-} \frac { 1 } { 4 } & - \frac { 1 } { 4 } & \phantom{-} \frac { 1 } { 4 } & \phantom{-} \frac { 1 } { 4 } &
			\phantom{-} \frac { 1 } { 4 } & - \frac { 1 } { 4 } & \phantom{-} \frac { 1 } { 4 } & \phantom{-} \frac { 1 } { 4 } &
			- \frac { 1 } { 4 } & \phantom{-} \frac { 1 } { 4 } & - \frac { 1 } { 4 } & - \frac { 1 } { 4 } &
			\phantom{-} \frac { 1 } { 4 } & - \frac { 1 } { 4 } & \phantom{-} \frac { 1 } { 4 } & \phantom{-} \frac { 1 } { 4 }
			\\
			\phantom{-} \frac { 1 } { 4 } & \phantom{-} \frac { 1 } { 4 } & - \frac { 1 } { 4 } & \phantom{-} \frac { 1 } { 4 } &
			\phantom{-} \frac { 1 } { 4 } & \phantom{-} \frac { 1 } { 4 } & - \frac { 1 } { 4 } & \phantom{-} \frac { 1 } { 4 } &
			- \frac { 1 } { 4 } & - \frac { 1 } { 4 } & \phantom{-} \frac { 1 } { 4 } & - \frac { 1 } { 4 } &
			\phantom{-} \frac { 1 } { 4 } & \phantom{-} \frac { 1 } { 4 } & - \frac { 1 } { 4 } & \phantom{-} \frac { 1 } { 4 }
			\\
			\phantom{-} \frac { 1 } { 4 } & \phantom{-} \frac { 1 } { 4 } & \phantom{-} \frac { 1 } { 4 } & - \frac { 1 } { 4 } &
			\phantom{-} \frac { 1 } { 4 } & \phantom{-} \frac { 1 } { 4 } & \phantom{-} \frac { 1 } { 4 } & - \frac { 1 } { 4 } &
			- \frac { 1 } { 4 } & - \frac { 1 } { 4 } & - \frac { 1 } { 4 } & \phantom{-} \frac { 1 } { 4 } &
			\phantom{-} \frac { 1 } { 4 } & \phantom{-} \frac { 1 } { 4 } & \phantom{-} \frac { 1 } { 4 } & - \frac { 1 } { 4 }
			\\
			- \frac { 1 } { 4 } & \phantom{-} \frac { 1 } { 4 } & \phantom{-} \frac { 1 } { 4 } & \phantom{-} \frac { 1 } { 4 } &
			- \frac { 1 } { 4 } & \phantom{-} \frac { 1 } { 4 } & \phantom{-} \frac { 1 } { 4 } & \phantom{-} \frac { 1 } { 4 } &
			- \frac { 1 } { 4 } & \phantom{-} \frac { 1 } { 4 } & \phantom{-} \frac { 1 } { 4 } & \phantom{-} \frac { 1 } { 4 } &
			\phantom{-} \frac { 1 } { 4 } & - \frac { 1 } { 4 } & - \frac { 1 } { 4 } & - \frac { 1 } { 4 }
			\\
			\phantom{-} \frac { 1 } { 4 } & - \frac { 1 } { 4 } & \phantom{-} \frac { 1 } { 4 } & \phantom{-} \frac { 1 } { 4 } &
			\phantom{-} \frac { 1 } { 4 } & - \frac { 1 } { 4 } & \phantom{-} \frac { 1 } { 4 } & \phantom{-} \frac { 1 } { 4 } &
			\phantom{-} \frac { 1 } { 4 } & - \frac { 1 } { 4 } & \phantom{-} \frac { 1 } { 4 } & \phantom{-} \frac { 1 } { 4 } &
			- \frac { 1 } { 4 } & \phantom{-} \frac { 1 } { 4 } & - \frac { 1 } { 4 } & - \frac { 1 } { 4 }
			\\
			\phantom{-} \frac { 1 } { 4 } & \phantom{-} \frac { 1 } { 4 } & - \frac { 1 } { 4 } & \phantom{-} \frac { 1 } { 4 } &
			\phantom{-} \frac { 1 } { 4 } & \phantom{-} \frac { 1 } { 4 } & - \frac { 1 } { 4 } & \phantom{-} \frac { 1 } { 4 } &
			\phantom{-} \frac { 1 } { 4 } & \phantom{-} \frac { 1 } { 4 } & - \frac { 1 } { 4 } & \phantom{-} \frac { 1 } { 4 } &
			- \frac { 1 } { 4 } & - \frac { 1 } { 4 } & \phantom{-} \frac { 1 } { 4 } & - \frac { 1 } { 4 }
			\\
			\phantom{-} \frac { 1 } { 4 } & \phantom{-} \frac { 1 } { 4 } & \phantom{-} \frac { 1 } { 4 } & - \frac { 1 } { 4 } &
			\phantom{-} \frac { 1 } { 4 } & \phantom{-} \frac { 1 } { 4 } & \phantom{-} \frac { 1 } { 4 } & - \frac { 1 } { 4 } &
			\phantom{-} \frac { 1 } { 4 } & \phantom{-} \frac { 1 } { 4 } & \phantom{-} \frac { 1 } { 4 } & - \frac { 1 } { 4 } &
			- \frac { 1 } { 4 } & - \frac { 1 } { 4 } & - \frac { 1 } { 4 } & \phantom{-} \frac { 1 } { 4 }
			\CodeAfter
			\tikz \draw
				[
					purple4,
					line width = 0.100 mm,
					shorten <= 2.000 mm,
					shorten >= 2.000 mm,
				]
			(9 -| 1) -- (9 -| last);
			\tikz \draw
				[
					purple4,
					line width = 0.100 mm,
					shorten <= 2.000 mm,
				]
				(1 -| 9) -- (last -| 9);
		\end{bNiceMatrix}
	}
	\ .
\end{align}

\end{example}

We use the unitary classifier $C_{ I_{ 2 n } }$, $n \geq 1$, as the main component in the construction of the family of quantum circuits QCPC$_{ 2 n }$, for the classification of the class of Boolean functions $F_{ 2 n }$. Suppose we intend to classify a Boolean function belonging to the $\mathcal{ F }$ sequence, and, in particular, to the class $F_{ 2 n }$, for some specific $n$. This is easily achievable by the quantum circuit QCPC$_{ 2 n }$ that utilizes as the primary component the unitary classifier $C_{ I_{ 2 n } }$. The situation is summarized in the following Theorem \ref{thr: Classification In The Promised Class} from \cite{Andronikos2025a}.

\begin{theorem} {Classification in the Promised Class} { Classification In The Promised Class}
	Let $f_{ i }$, $0 \leq i \leq 2^{ 2 n } - 1$, be a Boolean function contained in the class $F_{ 2 n }$ $=$ $( f_{ 0 }, f_{ 1 },$ \dots $, f_{ 2^{ 2 n } - 1 } )$. Upon measuring the quantum circuit QCPC$_{ 2 n }$ containing the oracle for $f_{ i }$ and the unitary classifier $C_{ I_{ 2 n } }$, we obtain $\ket{ \mathbf{ i } }$, where $\mathbf{ i }$ is the binary representation of the index $i$, and $\ket{ \mathbf{ i } }$ is one of the basis kets of the computational basis.
\end{theorem}

\begin{example} {Classifying functions in the Promised Class} { Classifying Functions In The Promised Class}

Suppose that, in the Classification Game between Bob and Alice, Bob chooses a Boolean function from $F_{ 4 }$, which belongs to the promised class of functions. Say that Bob chooses $f_{ 3 }$, the behavior of which is encoded by the pattern vector $\mathbf{ p }_{ 3 } = 1000 \ 1000 \ 1000 \ 0111$, listed in Example \ref{xmp: Product Examples}. Alice makes her move by employing the classifier $C_{ I_{ 4 } } = C_{ I_{ 2 } }^{ \otimes 2 }$. In this case, the concrete implementation in Qiskit \cite{Qiskit2025} of the classification algorithm takes the form shown in Figure \ref{fig: Phase4___1000100010000111___}, where Bob uses the oracle for the function $f_{ 3 }$ and Alice the classifier $C_{ I_{ 4 } }$.

\begin{tcolorbox}
	[
		enhanced,
		breakable,
		center title,
		fonttitle = \bfseries,
		grow to left by = 0.000 cm,
		grow to right by = 0.000 cm,
		colback = white,						
		enhanced jigsaw,						
		sharp corners,
		toprule = 0.001 pt,
		bottomrule = 0.001 pt,
		leftrule = 0.001 pt,
		rightrule = 0.001 pt,
	]
	\begin{figure}[H]
			\centering
			\includegraphics [ scale = 0.515, trim = {0.850cm 0.000cm 0.000cm 0.750cm}, clip ] {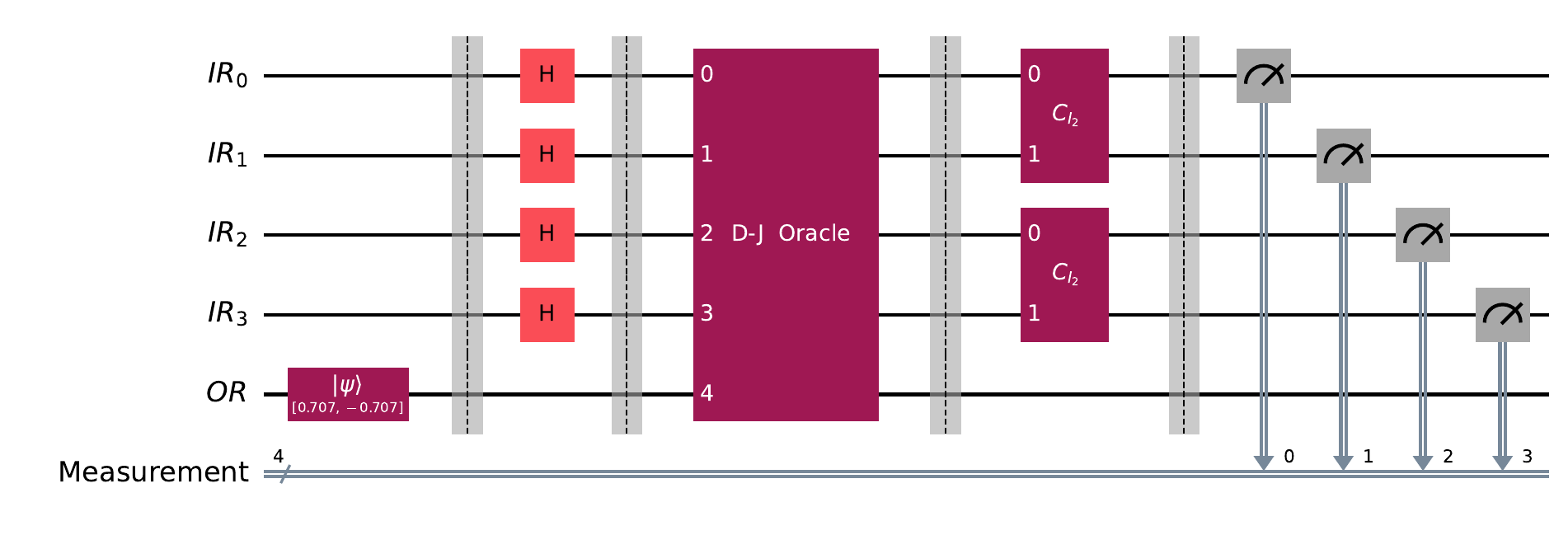}
			\caption{This figure shows the implementation of the BFPQC algorithm for the classification of the Boolean functions in $F_{ 4 }$, assuming Bob has chosen the oracle for the function $f_{ 3 }$ and Alice has employed the classifier $C_{ I_{ 4 } }$.}
			\label{fig: Phase4___1000100010000111___}
		\end{figure}
\end{tcolorbox}

\begin{tcolorbox}
	[
		enhanced,
		breakable,
		center title,
		fonttitle = \bfseries,
		grow to left by = 0.000 cm,
		grow to right by = 0.000 cm,
		colback = white,						
		enhanced jigsaw,							
		sharp corners,
		toprule = 0.001 pt,
		bottomrule = 0.001 pt,
		leftrule = 0.001 pt,
		rightrule = 0.001 pt,
	]
	\begin{minipage} [ b ] { 0.450 \textwidth }
		After the action of the classifier, the state of the system is just $\ket{ \mathbf{ 0011 } }$. Therefore, measuring the quantum circuit depicted in Figure \ref{fig: Phase4___1000100010000111___} will output the bit vector $0011$ with probability $1$ (as corroborated by the measurements contained in Figure \ref{fig: Phase4_Histogram_StatevectorSampler___1000100010000111___}), which is the binary representation of the index of $f_{ 3 }$. Alice surely wind the game, as anticipated.
	\end{minipage}
	\hspace{ 0.300 cm }
	\begin{minipage} [ b ] { 0.450 \textwidth }
		\begin{figure}[H]
			\centering
			\includegraphics [ scale = 0.500, trim = {0.000cm 0.000cm 0.000cm 0.000cm}, clip ] {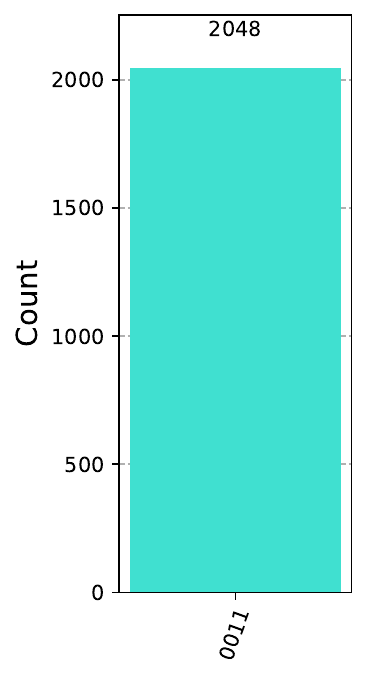}
			\caption{The measurement outcome of the quantum circuit of Figure \ref{fig: Phase4___1000100010000111___}.}
			\label{fig: Phase4_Histogram_StatevectorSampler___1000100010000111___}
		\end{figure}
	\end{minipage}
\end{tcolorbox}

\end{example}

Let us assume that we are given the quantum circuit visualized in Figure \ref{fig: The BFPQC Quantum Circuit for $F_{ 2 n }$}. If the oracle encodes a function of $F_{ 2 n }$, this circuit will conclusively, i.e., with probability $1.0$, determine this function.
When the oracle realizes a function outside $F_{ 2 n }$, the final measurement will not reveal the function. However, as we shall demonstrate, it is possible to obtain some useful information when the input function is sufficiently close to the class $F_{ 2 n }$, in the sense of Definition \ref{def: Left & Right Clusters}.

\begin{tcolorbox}
	[
		enhanced,
		breakable,
		center title,
		fonttitle = \bfseries,
		grow to left by = 0.000 cm,
		grow to right by = 0.000 cm,
		colback = white,						
		enhanced jigsaw,							
		sharp corners,
		toprule = 0.001 pt,
		bottomrule = 0.001 pt,
		leftrule = 0.001 pt,
		rightrule = 0.001 pt,
	]
	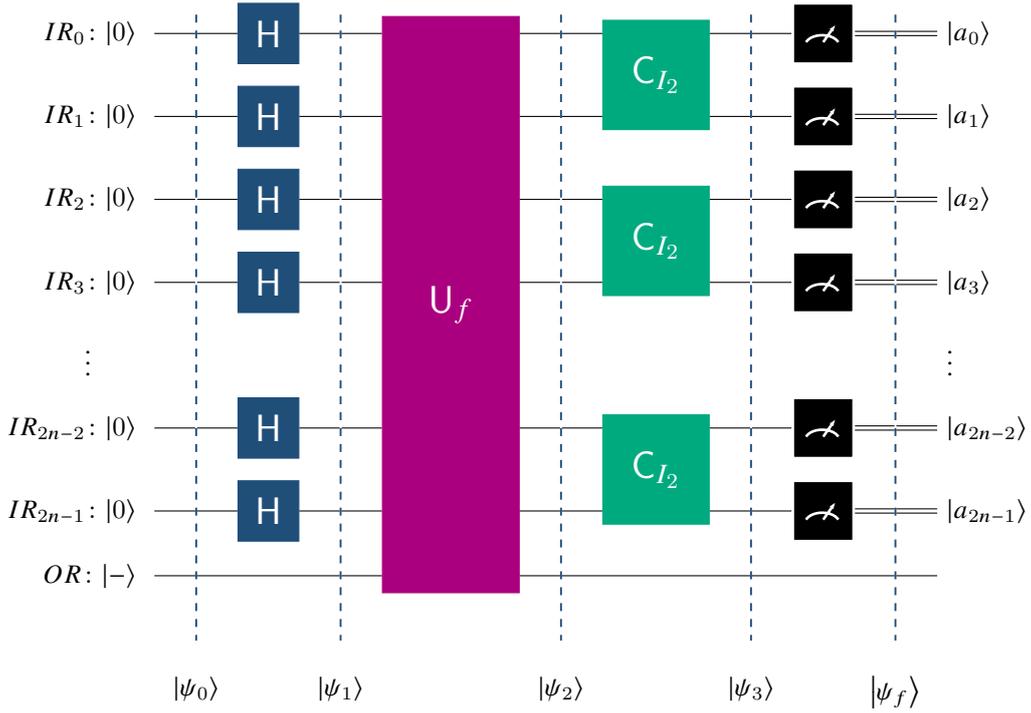
\begin{figure}[H]
		\centering
		\begin{tikzpicture} [ scale = 1.000 ] 
			\begin{yquant}[ operator/separation = 3.000 mm, register/separation = 3.000 mm, every nobit output/.style = { } ]
				qubit { $IR_{ 0 } \colon \ket{ 0 }$ } IR;
				qubit { $IR_{ 1 } \colon \ket{ 0 }$ } IR [ + 1 ];
				qubit { $IR_{ 2 } \colon \ket{ 0 }$ } IR [ + 1 ];
				qubit { $IR_{ 3 } \colon \ket{ 0 }$ } IR [ + 1 ];
				qubit { $\vdots$ \hspace{ 0.475 cm } } IR [ + 1 ]; discard IR [ 4 ];
				qubit { $IR_{ 2 n - 2 } \colon \ket{ 0 }$ } IR [ + 1 ];
				qubit { $IR_{ 2 n - 1 } \colon \ket{ 0 }$ } IR [ + 1 ];
				qubit { $OR \colon \ket{ - }$ } OR;
				nobit AUX_1;
				[
				name = Input,
				WordBlueDarker,
				line width = 0.250 mm,
				]
				barrier ( - ) ;
				[ draw = WordBlueDarker, fill = WordBlueDarker, radius = 0.400 cm ] box {\color{white} \Large \sf{H}} IR [ 0 ];
				[ draw = WordBlueDarker, fill = WordBlueDarker, radius = 0.400 cm ] box {\color{white} \Large \sf{H}} IR [ 1 ];
				[ draw = WordBlueDarker, fill = WordBlueDarker, radius = 0.400 cm ] box {\color{white} \Large \sf{H}} IR [ 2 ];
				[ draw = WordBlueDarker, fill = WordBlueDarker, radius = 0.400 cm ] box {\color{white} \Large \sf{H}} IR [ 3 ];
				[ draw = WordBlueDarker, fill = WordBlueDarker, radius = 0.400 cm ] box {\color{white} \Large \sf{H}} IR [ 5 ];
				[ draw = WordBlueDarker, fill = WordBlueDarker, radius = 0.400 cm ] box {\color{white} \Large \sf{H}} IR [ 6 ];
				[
				name = Expansion,
				WordBlueDarker,
				line width = 0.250 mm,
				]
				barrier ( - ) ;
				[ draw = RedPurple, fill = RedPurple, x radius = 0.900 cm, y radius = 0.450 cm ] box { \color{white} \Large \sf{U}$_{ f }$} ( IR - OR );
				[
				name = Oracle,
				WordBlueDarker,
				line width = 0.250 mm,
				]
				barrier ( - ) ;
				[ draw = GreenLighter2, fill = GreenLighter2, x radius = 0.700 cm, y radius = 0.350 cm ] box { \color{white} \Large \sf{C}$_{ I_{ 2 } }$}  ( IR [ 0 ] - IR [ 1 ] );
				[ draw = GreenLighter2, fill = GreenLighter2, x radius = 0.700 cm, y radius = 0.350 cm ] box { \color{white} \Large \sf{C}$_{ I_{ 2 } }$}  ( IR [ 2 ] - IR [ 3 ] );
				[ draw = GreenLighter2, fill = GreenLighter2, x radius = 0.700 cm, y radius = 0.350 cm ] box { \color{white} \Large \sf{C}$_{ I_{ 2 } }$}  ( IR [ 5 ] - IR [ 6 ] );
				[
				name = Classifier,
				WordBlueDarker,
				line width = 0.250 mm,
				]
				barrier ( - ) ;
				[ line width = .350 mm, draw = white, fill = black, radius = 0.400 cm ] measure IR [ 0 ];
				[ line width = .350 mm, draw = white, fill = black, radius = 0.400 cm ] measure IR [ 1 ];
				[ line width = .350 mm, draw = white, fill = black, radius = 0.400 cm ] measure IR [ 2 ];
				[ line width = .350 mm, draw = white, fill = black, radius = 0.400 cm ] measure IR [ 3 ];
				[ line width = .350 mm, draw = white, fill = black, radius = 0.400 cm ] measure IR [ 5 ];
				[ line width = .350 mm, draw = white, fill = black, radius = 0.400 cm ] measure IR [ 6 ];
				[
				name = Measurement,
				WordBlueDarker,
				line width = 0.250 mm,
				]
				barrier ( - ) ;
				output { $\ket{ a_{ 0 } }$ } IR [ 0 ];
				output { $\ket{ a_{ 1 } }$ } IR [ 1 ];
				output { $\ket{ a_{ 2 } }$ } IR [ 2 ];
				output { $\ket{ a_{ 3 } }$ } IR [ 3 ];
				output { $\vdots$ } IR [ 4 ];
				output { $\ket{ a_{ 2 n - 2 } }$ } IR [ 5 ];
				output { $\ket{ a_{ 2 n - 1 } }$ } IR [ 6 ];
				\node [ below = 4.500 cm ] at (Input) { $\ket{ \psi_{ 0 } }$ };
				\node [ below = 4.500 cm ] at (Expansion) { $\ket{ \psi_{ 1 } }$ };
				\node [ below = 4.500 cm ] at (Oracle) { $\ket{ \psi_{ 2 } }$ };
				\node [ below = 4.500 cm ] at (Classifier) { $\ket{ \psi_{ 3 } }$ };
				\node [ below = 4.500 cm ] at (Measurement) { $\ket{ \psi_{ f } }$ };
			\end{yquant}
		\end{tikzpicture}
		\caption{This figure depicts the quantum circuit that conclusively classifies all functions contained in $F_{ 2 n }$.}
		\label{fig: The BFPQC Quantum Circuit for $F_{ 2 n }$}
	\end{figure}
\end{tcolorbox}

To enhance clarity, we explain the notation used in the above figure.

\begin{itemize}
	\item	
	$IR$ is the quantum input register that contains $2 n$ qubits and starts its operation at state $\ket{ \mathbf{ 0 } }$.
	\item	
	$OR$ is the single-qubit output register initialized to $\ket{ - }$.
	\item	
	$H$ is the Hadamard transform.
	\item	
	$U_{ f }$ is the unitary transform corresponding to the oracle for the unknown function $h$.
	\item	
	$C_{ I_{ 2 } }$ is the fundamental building block of $C_{ I_{ 2 n } }$, as evidenced by equation \eqref{eq: Unitary Pattern Classifier $C_{ I_{ 2 ( n + 1 ) } }$}.
\end{itemize}
\section{Outside the promised class} \label{sec: Outside the Promised Class}

For our investigation we shall consider classes of balanced Boolean functions. We begin by first defining the next sequence of balanced pattern bases, as outlined in the next Definition \ref{def: A Sequence Of Balanced Pattern Bases}. To avoid any confusion, we clarify that only the first pattern bit vector in every basis represents the constant function; all the remaining pattern bit vectors encode balanced functions.

\begin{definition} {A Sequence of Balanced Pattern Bases} { A Sequence Of Balanced Pattern Bases}
	We recursively define a sequence $( B_{ 2 m } )_{ m \geq 1 }$, of pattern bases demonstrating balanced behavior.
	\begin{enumerate}
		[ left = 0.600 cm, labelsep = 0.500 cm, start = 0 ]
		\renewcommand \labelenumi { $($\textbf{B}$_{ \theenumi }$$)$ }
		\item	If $m = 1$, the corresponding pattern basis is $B_{ 2 }$, as defined below
		\begin{align}
			\label{eq: Pattern Basis $B_{ 2 }$}
			&B_{ 2 }
			\coloneq
			(
			\mathbf{ e }_{ 0 },
			\mathbf{ e }_{ 1 },
			\mathbf{ e }_{ 2 },
			\mathbf{ e }_{ 3 }
			)
			\ ,
			\nonumber
			\\
			&\text{where}
			\nonumber
			\\
			&\mathbf{ e }_{ 0 } = 0000,
			\mathbf{ e }_{ 1 } = 0101,
			\mathbf{ e }_{ 2 } = 0011,
			\mathbf{ e }_{ 3 } = 0110
			\ .
		\end{align}
		\item	Given the pattern basis $B_{ 2 m }$ of rank $2 m$, we define the pattern basis $B_{ 2 ( m + 1 ) }$ of rank $2 ( m + 1 )$ as follows:
		\begin{align}
			\label{eq: Pattern Basis $B_{ 2 ( m + 1 ) }$}
			B_{ 2 ( m + 1 ) }
			\coloneq
			B_{ 2 }
			\odot
			B_{ 2 m }
			\ .
		\end{align}
		\item	This sequence of pattern bases $B_{ 2 }$, $B_{ 4 }$, \dots, $B_{ 2 m }$, \dots, is denoted by $\mathcal{ B }$.
	\end{enumerate}
\end{definition}

\begin{definition} {Functions from Balanced Bases} { Functions From Balanced Bases}
	Let $B_{ 2 m }$ $=$ $( \mathbf{ e }_{ 0 }, \mathbf{ e }_{ 1 },$ \dots $, \mathbf{ e }_{ 2^{ 2 m } - 1 } )$ be a balanced pattern basis of rank $2 m$. To $B_{ 2 m }$ we associate the class $G_{ 2 m }$ $=$ $( g_{ 0 }, g_{ 1 },$ \dots $, g_{ 2^{ 2 m } - 1 } )$ of Boolean functions from $\mathbb{ B }^{ 2 m }$ to $\mathbb{ B }$ such that $\mathbf{ e }_{ i }$ is the pattern bit vector corresponding to $g_{ i }$, $0 \leq i \leq 2^{ 2 m } - 1$. This sequence of classes of Boolean functions $G_{ 2 }$, $G_{ 4 }$, \dots, $G_{ 2 m }$, \dots, is denoted by $\mathcal{ G }$.
\end{definition}

It is immediately evident that the pattern bit vectors in $B_{ 2 }$ correspond to binary functions outside $F_{ 2 }$. These particular pattern bit vectors are inspired from the $2$-fold Hadamard transform $H^{ \otimes 2 }$. Given any $\mathbf{ e }_{ i } \in B_{ 2 }$, $0 \leq i \leq 3$, the corresponding function can't be classified by BFPQC. It's very easy to see why this holds. The pattern kets corresponding to the bit vectors of $B_{ 2 }$ are

\begin{align*}
	\ket{ a_{ \mathbf{ e_{ 0 } } } }
	&=
	\frac { 1 } { 2 } \ket{ 00 } +
	\frac { 1 } { 2 } \ket{ 01 } +
	\frac { 1 } { 2 } \ket{ 10 } +
	\frac { 1 } { 2 } \ket{ 11 }
	&
	\ket{ a_{ \mathbf{ e_{ 1 } } } }
	&=
	\frac { 1 } { 2 } \ket{ 00 } -
	\frac { 1 } { 2 } \ket{ 01 } +
	\frac { 1 } { 2 } \ket{ 10 } -
	\frac { 1 } { 2 } \ket{ 11 }
	\\
	\ket{ a_{ \mathbf{ e_{ 2 } } } }
	&=
	\frac { 1 } { 2 } \ket{ 00 } +
	\frac { 1 } { 2 } \ket{ 01 } -
	\frac { 1 } { 2 } \ket{ 10 } -
	\frac { 1 } { 2 } \ket{ 11 }
	&
	\ket{ a_{ \mathbf{ e_{ 3 } } } }
	&=
	\frac { 1 } { 2 } \ket{ 00 } -
	\frac { 1 } { 2 } \ket{ 01 } -
	\frac { 1 } { 2 } \ket{ 10 } +
	\frac { 1 } { 2 } \ket{ 11 }
	\ ,
\end{align*}

which are precisely the columns of $H^{ \otimes 2 }$. Accordingly, the action of the $C_{ I_{ 2 } }$ classifier on any of them will result in one of the following states

\begin{align*}
	C_{ I_{ 2 } }
	\ket{ a_{ \mathbf{ e_{ 0 } } } }
	&=
	\frac { 1 } { 2 } \ket{ 00 } +
	\frac { 1 } { 2 } \ket{ 01 } +
	\frac { 1 } { 2 } \ket{ 10 } +
	\frac { 1 } { 2 } \ket{ 11 }
	&
	C_{ I_{ 2 } }
	\ket{ a_{ \mathbf{ e_{ 1 } } } }
	&=
	- \frac { 1 } { 2 } \ket{ 00 } +
	\frac { 1 } { 2 } \ket{ 01 } -
	\frac { 1 } { 2 } \ket{ 10 } +
	\frac { 1 } { 2 } \ket{ 11 }
	\\
	C_{ I_{ 2 } }
	\ket{ a_{ \mathbf{ e_{ 2 } } } }
	&=
	- \frac { 1 } { 2 } \ket{ 00 } -
	\frac { 1 } { 2 } \ket{ 01 } +
	\frac { 1 } { 2 } \ket{ 10 } +
	\frac { 1 } { 2 } \ket{ 11 }
	&
	C_{ I_{ 2 } }
	\ket{ a_{ \mathbf{ e_{ 3 } } } }
	&=
	- \frac { 1 } { 2 } \ket{ 00 } +
	\frac { 1 } { 2 } \ket{ 01 } +
	\frac { 1 } { 2 } \ket{ 10 } -
	\frac { 1 } { 2 } \ket{ 11 }
	\ ,
\end{align*}

which, upon measurement, will collapse of any of the four basis states with equal probability $\frac { 1 } { 4 }$. Reasoning in the same manner and using induction, we immediately arrive at the next Theorem \ref{thr: Functions From The G Sequence}.

\begin{theorem} {Functions Outside the Class} { Functions From The G Sequence}
	Let $g_{ i }$, $0 \leq i \leq 2^{ 2 m } - 1$, be a Boolean function contained in the class $G_{ 2 n }$ $=$ $( g_{ 0 }, g_{ 1 },$ \dots $, g_{ 2^{ 2 m } - 1 } )$. The state of the input register $IR$ in the quantum circuit QCPC$_{ 2 m }$ that contains the oracle for $g_{ i }$ and the unitary classifier $C_{ I_{ 2 m } }$, before measurement is
	\begin{align}
		\label{eq: QCPC$_{ 2 m }$ State Before Measurement}
		2^{ - m }
		\sum_{ \mathbf{ z } \in \mathbb{ B }^{ 2 m } }
		( - 1 )^{ \mathbf{ z } \cdot \mathbf{ i } }
		\ket{ \mathbf{ z } }
		\ .
	\end{align}
	Thus, upon measurement  of the input register $IR$, we may obtain any of the $2^{ 2 m }$ basis kets of the computational basis with equal probability $2^{ - 2 m }$, rendering classification probabilistically impossible.
\end{theorem}

\begin{definition} {Extended Product} { Extended Product}
	Let $f \colon \mathbb{ B }^{ n } \rightarrow \mathbb{ B }$ and $g \colon \mathbb{ B }^{ m } \rightarrow \mathbb{ B }$ be two Boolean functions. We construct their \emph{extended product}, denoted by $h = f \star g$, as the Boolean function $h \colon \mathbb{ B }^{ n + m } \rightarrow \mathbb{ B }$ defined as follows:
	\begin{align}
		\label{eq: Extended Product}
		h ( j + i 2^{ n }  )
		=
		\begin{cases}
			\ g ( j )						& \text{if } f ( i ) = 0 
			\\[ 1.000 ex ]
			\ \overline{ g ( j ) }			& \text{if } f ( i ) = 1 
		\end{cases}
		\ ,
		\quad
		0 \leq i \leq 2^{ n } - 1,
		\quad
		0 \leq j \leq 2^{ m } - 1
		\ .
	\end{align}
\end{definition}

\begin{proposition} {Equivalence of Products} { Equivalence Of Products}
	Let $h$ be the extended product of $f$ and $g$, let $\mathbf{ p }_{ h }$, $\mathbf{ p }_{ f }$ and $\mathbf{ p }_{ g }$ be their corresponding pattern bit vectors, and let $\ket* { a_{ \mathbf{ p }_{ h } } }$, $\ket* { a_{ \mathbf{ p }_{ f } } }$ and $\ket* { a_{ \mathbf{ p }_{ g } } }$ be their corresponding pattern kets. Then the following relations hold:
	\begin{align}
		\label{eq: Equivalence Of Products}
		h
		=
		f
		\star
		g
		\Leftrightarrow
		\mathbf{ p }_{ h }
		=
		\mathbf{ p }_{ f }
		\odot
		\mathbf{ p }_{ g }
		\Leftrightarrow
		\ket* { a_{ \mathbf{ p }_{ h } } }
		=
		\ket* { a_{ \mathbf{ p }_{ f } } }
		\otimes
		\ket* { a_{ \mathbf{ p }_{ g } } }
		\ .
	\end{align}
\end{proposition}

\begin{example} {Extended Product Example} { Extended Product Example}

	This example explains how the extended product between Boolean functions works. Consider two functions from the class $F_{ 2 }$, say $f_{ 0 }$ and $f_{ 3 }$, corresponding to pattern bit vectors $\mathbf{ p }_{ 0 }$ and $\mathbf{ p }_{ 3 }$ from \eqref{eq: Pattern Basis I_2}.

\begin{table}[H]
	\caption{The truth table of the function $f_{ 0 } \star f_{ 3 }$.}
	\label{tbl: Truth Table Of The function f}
	\centering
	\SetTblrInner { rowsep = 1.000 mm }
	\begin{tblr}
		{
			colspec =
			{
				Q [ c, m, 0.500 cm ]
				| [ 0.500 pt, DeepPink4 ]
				Q [ c, m, 1.500 cm ]
				| [ 0.500 pt, DeepPink4 ]
				Q [ c, m, 0.500 cm ]
				| [ 0.500 pt, DeepPink4 ]
				Q [ c, m, 1.500 cm ]
				| [ 0.500 pt, DeepPink4 ]
				Q [ c, m, 1.500 cm ]
				| [ 0.500 pt, DeepPink4 ]
				| [ 0.500 pt, DeepPink4 ]
				Q [ c, m, 0.500 cm ]
				| [ 1.500 pt, DeepPink4 ]
				Q [ c, m, 2.250 cm ]
			},
			rowspec =
			{
				|
				[ 3.500 pt, DeepPink4 ]
				|
				[ 0.750 pt, DeepPink4 ]
				|
				[ 0.250 pt, white ]
				Q
				|
				[ 0.750 pt, DeepPink4 ]
				Q
				|
				[ 0.100 pt, DeepPink4 ]
				Q
				|
				[ 0.100 pt, DeepPink4 ]
				Q
				|
				[ 0.100 pt, DeepPink4 ]
				Q
				|
				[ 0.750 pt, DeepPink4 ]
				Q
				|
				[ 0.100 pt, DeepPink4 ]
				Q
				|
				[ 0.100 pt, DeepPink4 ]
				Q
				|
				[ 0.100 pt, DeepPink4 ]
				Q
				|
				[ 0.750 pt, DeepPink4 ]
				Q
				|
				[ 0.100 pt, DeepPink4 ]
				Q
				|
				[ 0.100 pt, DeepPink4 ]
				Q
				|
				[ 0.100 pt, DeepPink4 ]
				Q
				|
				[ 0.750 pt, DeepPink4 ]
				Q
				|
				[ 0.100 pt, DeepPink4 ]
				Q
				|
				[ 0.100 pt, DeepPink4 ]
				Q
				|
				[ 0.100 pt, DeepPink4 ]
				Q
				|
				[ 3.500 pt, DeepPink4 ]
			}
		}
		\SetCell { bg = VioletRed4, fg = white } $i$
		&
		$f_{ 0 } ( x_{ 1 }, x_{ 0 } )$
		&
		\SetCell { bg = VioletRed4, fg = white } $j$
		&
		$f_{ 3 } ( x_{ 1 }, x_{ 0 } )$
		&
		$\overline{ f_{ 3 } ( x_{ 1 }, x_{ 0 } ) }$
		&
		\SetCell { bg = VioletRed4, fg = white } $k$
		&
		$f ( x_{ 3 }, x_{ 2 }, x_{ 1 }, x_{ 0 } )$
		\\
		$0$ & $1$ & $0$ & $0$ & $1$ & $0$ & \SetCell { bg = VioletRed4, fg = white } $1$
		\\
		$0$ & $1$ & $0$ & $0$ & $1$ & $1$ & \SetCell { bg = VioletRed4, fg = white } $1$
		\\
		$0$ & $1$ & $0$ & $0$ & $1$ & $2$ & \SetCell { bg = VioletRed4, fg = white } $1$
		\\
		$0$ & $1$ & $0$ & $1$ & $0$ & $3$ & \SetCell { bg = VioletRed4, fg = white } $0$
		\\
		$1$ & $0$ & $1$ & $0$ & $1$ & $4$ & \SetCell { bg = VioletRed4, fg = white } $0$
		\\
		$1$ & $0$ & $1$ & $0$ & $1$ & $5$ & \SetCell { bg = VioletRed4, fg = white } $0$
		\\
		$1$ & $0$ & $1$ & $0$ & $1$ & $6$ & \SetCell { bg = VioletRed4, fg = white } $0$
		\\
		$1$ & $0$ & $1$ & $1$ & $0$ & $7$ & \SetCell { bg = VioletRed4, fg = white } $1$
		\\
		$2$ & $0$ & $2$ & $0$ & $1$ & $8$ & \SetCell { bg = VioletRed4, fg = white } $0$
		\\
		$2$ & $0$ & $2$ & $0$ & $1$ & $9$ & \SetCell { bg = VioletRed4, fg = white } $0$
		\\
		$2$ & $0$ & $2$ & $0$ & $1$ & $10$ & \SetCell { bg = VioletRed4, fg = white } $0$
		\\
		$2$ & $0$ & $2$ & $1$ & $0$ & $11$ & \SetCell { bg = VioletRed4, fg = white } $1$
		\\
		$3$ & $0$ & $3$ & $0$ & $1$ & $12$ & \SetCell { bg = VioletRed4, fg = white } $0$
		\\
		$3$ & $0$ & $3$ & $0$ & $1$ & $13$ & \SetCell { bg = VioletRed4, fg = white } $0$
		\\
		$3$ & $0$ & $3$ & $0$ & $1$ & $14$ & \SetCell { bg = VioletRed4, fg = white } $0$
		\\
		$3$ & $0$ & $3$ & $1$ & $0$ & $15$ & \SetCell { bg = VioletRed4, fg = white } $1$
		\\
	\end{tblr}
\end{table}

\begin{align*}
	f_{ 0 }
	(
	x_{ 1 },
	x_{ 0 }
	)
	&\coloneq
	\neg x_{ 1 }
	\wedge
	\neg x_{ 0 }
	\\
	f_{ 3 }
	(
	x_{ 1 },
	x_{ 0 }
	)
	&\coloneq
	x_{ 1 }
	\wedge
	x_{ 0 }
\end{align*}

The detailed constructed of their extended product function $f_{ 0 } \star f_{ 3 }$ is contained in Table \ref{tbl: Truth Table Of The function f}. The pattern bit vectors and the pattern kets of $f_{ 0 }$, $f_{ 3 }$ and $f_{ 0 } \star f_{ 3 }$ are shown in Table \ref{tbl: Pattern Bit Vectors & Pattern Kets Products}, where the validity of Proposition \ref{prp: Equivalence Of Products} can be immediately verified. The resulting function belongs to the class $F_{ 4 }$, and the corresponding pattern bit vector is listed in Table \ref{tbl: Pattern Basis $I_{ 4 }$}, as $\mathbf{ r }_{ 3 }$.

\begin{table}[H]
	\caption{This table contains the pattern bit vectors and the pattern kets of $f_{ 0 }$, $f_{ 3 }$ and $f_{ 0 } \star f_{ 3 }$.}
	\label{tbl: Pattern Bit Vectors & Pattern Kets Products}
	\centering
	\SetTblrInner { rowsep = 1.200 mm }
	\begin{tblr}
		{
			colspec =
			{
				Q [ c, m, 2.000 cm ]
				| [ 0.750 pt, azure3 ]
				| [ 0.750 pt, azure3 ]
				Q [ c, m, 3.000 cm ]
				| [ 0.500 pt, azure3 ]
				| [ 0.500 pt, azure3 ]
				Q [ c, m, 7.000 cm ]
			},
			rowspec =
			{
				| [ 3.500 pt, azure3 ]
				| [ 0.750 pt, azure3 ]
				| [ 0.250 pt, white ]
				Q
				|
				Q
				| [ 0.150 pt, azure3 ]
				Q
				| [ 0.150 pt, azure3 ]
				Q
				| [ 3.500 pt, azure3 ]
			}
		}
		\SetCell { bg = azure4, fg = white } Function
		&
		\SetCell { bg = azure5, fg = white } Pattern Bit Vector
		&
		\SetCell { bg = azure6, fg = white } Pattern Ket
		\\
		$f_{ 0 }$
		&
		$0001$
		&
		$- \frac { 1 } { 2 } \ket{ 00 } + \frac { 1 } { 2 } \ket{ 01 } + \frac { 1 } { 2 } \ket{ 10 } + \frac { 1 } { 2 } \ket{ 11 }$
		\\
		$f_{ 3 }$
		&
		$1000$
		&
		$\frac { 1 } { 2 } \ket{ 00 } + \frac { 1 } { 2 } \ket{ 01 } + \frac { 1 } { 2 } \ket{ 10 } - \frac { 1 } { 2 } \ket{ 11 }$
		\\
		$f_{ 0 } \star f_{ 3 }$
		&
		$1000 \ 1000 \ 1000 \ 0111$
		&
		\begin{align*}
			- \frac { 1 } { 4 } \ket{ 0000 } - \frac { 1 } { 4 } \ket{ 0001 } - \frac { 1 } { 4 } \ket{ 0010 } + \frac { 1 } { 4 } \ket{ 0011 }
			\\
			+ \frac { 1 } { 4 } \ket{ 0100 } + \frac { 1 } { 4 } \ket{ 0101 } + \frac { 1 } { 4 } \ket{ 0110 } - \frac { 1 } { 4 } \ket{ 0111 }
			\\
			+ \frac { 1 } { 4 } \ket{ 1000 } + \frac { 1 } { 4 } \ket{ 1001 } + \frac { 1 } { 4 } \ket{ 1010 } - \frac { 1 } { 4 } \ket{ 1011 }
			\\
			+ \frac { 1 } { 4 } \ket{ 1100 } + \frac { 1 } { 4 } \ket{ 1101 } + \frac { 1 } { 4 } \ket{ 1110 } - \frac { 1 } { 4 } \ket{ 1111 }
		\end{align*}
		\\
	\end{tblr}
\end{table}

\end{example}

As we have previously explained, our aim is to study what information can be obtained by the BFPQC algorithm if the input function is outside the promised class, but, at the same time ``close'' to some function of the $\mathcal{ F }$ sequence. The next Definition \ref{def: Left & Right Clusters} formalizes this concept of nearness.

\begin{definition} {Left \& Right Clusters} { Left & Right Clusters}
	Let $F_{ 2 n }$ and $G_{ 2 m }$, $n, m \geq 1$, be two classes of functions in the $\mathcal{ F }$ and $\mathcal{ G }$ sequences, respectively. We define the following two concepts.
	\begin{itemize}
		\item	
		The \emph{left cluster} of $F_{ 2 n }$ by $G_{ 2 m }$, denoted by $G_{ 2 m } \star F_{ 2 n }$, is the class of Boolean functions of the form $g \star f$, where $f \in F_{ 2 n }$ and $g \in G_{ 2 m }$.
		\item	
		The \emph{right cluster} of $F_{ 2 n }$ by $G_{ 2 m }$, denoted by $F_{ 2 n } \star G_{ 2 m }$, is the class of Boolean functions of the form $f \star g$, where $f \in F_{ 2 n }$ and $g \in G_{ 2 m }$.
	\end{itemize}
	Henceforth, we shall consider any function in $F_{ 2 n }$ as being \emph{close} to every function in the left or right cluster of $F_{ 2 n }$ by $G_{ 2 m }$.
\end{definition}

From now on, we assume that the oracle in Figure \ref{fig: The BFPQC Quantum Circuit for $F_{ 2 n }$} realizes a function $h$ that is either in the left, or right cluster of $F_{ 2 n }$. It is convenient to distinguish the following two cases, depending on whether the function $h$ belongs to the left or right cluster of $F_{ 2 n }$.

\begin{itemize}
	\item	
	Suppose that $h = g_{ j } \star f_{ i }$, where $g_{ j } \in G_{ 2 m }$ and $f_{ i } \in F_{ 2 n }$ with corresponding indices $i$ and $j$, respectively. The appropriate circuit in this scenario is depicted in Figure \ref{fig: The BFPQC Quantum Circuit for $g_{ j } star f_{ i }$}. In this case, we know by Proposition \ref{prp: Equivalence Of Products} that
	\begin{align}
		\label{eq: Left Cluster Pattern Kets}
		\ket* { a_{ \mathbf{ p }_{ h } } }
		=
		\ket* { a_{ \mathbf{ p }_{ g_{ j } } } }
		\otimes
		\ket* { a_{ \mathbf{ p }_{ f_{ i } } } }
		\ .
	\end{align}
	Combining equations \eqref{eq: QCPC$_{ 2 m }$ State Before Measurement} and \eqref{eq: Left Cluster Pattern Kets}, with Theorems \ref{thr: Classification In The Promised Class} and \ref{thr: Functions From The G Sequence}, we arrive the following relation
	\begin{align}
		\label{eq: Left Cluster Classification}
		C_{ I_{ 2 ( n + m ) } }
		\ket* { a_{ \mathbf{ p }_{ h } } }
		=
		C_{ I_{ 2 m } }
		\ket* { a_{ \mathbf{ p }_{ g_{ j } } } }
		\otimes
		C_{ I_{ 2 n } }
		\ket* { a_{ \mathbf{ p }_{ f_{ i } } } }
		=
		\left(
		2^{ - m }
		\sum_{ \mathbf{ z } \in \mathbb{ B }^{ 2 m } }
		( - 1 )^{ \mathbf{ z } \cdot \mathbf{ i } }
		\ket{ \mathbf{ z } }
		\right)
		\otimes
		\ket* { \mathbf{ i } }
		\ .
	\end{align}
	This is the state of the input register $IR$ of the system prior to the final measurement. Hence, upon measurement, the final outcome in the input register $IR$ will be $\mathbf{ x } \mathbf{ i }$, that is the $2 m$ most significant qubits will contain a random basis ket from $\mathbb{ B }^{ 2 m }$, and the $2 n$ least significant qubits will contain $\mathbf{ i }$, which correctly identifies the $f_{ i }$ function belonging to $F_{ 2 n }$.
	\item	
	Suppose that $h = f_{ i } \star g_{ j }$, where $f_{ i } \in F_{ 2 n }$ and $g_{ j } \in G_{ 2 m }$ with corresponding indices $i$ and $j$, respectively. This situation is visualized in Figure \ref{fig: The BFPQC Quantum Circuit for $f_{ i } star g_{ j }$}. Resorting to the same reasoning as in the previous case, we see that
	\begin{align}
		\label{eq: Right Cluster Pattern Kets}
		\ket* { a_{ \mathbf{ p }_{ h } } }
		=
		\ket* { a_{ \mathbf{ p }_{ f_{ i } } } }
		\otimes
		\ket* { a_{ \mathbf{ p }_{ g_{ j } } } }
		\ .
	\end{align}
	Combining equations \eqref{eq: QCPC$_{ 2 m }$ State Before Measurement} and \eqref{eq: Left Cluster Pattern Kets}, with Theorems \ref{thr: Classification In The Promised Class} and \ref{thr: Functions From The G Sequence}, we arrive the following relation
	\begin{align}
		\label{eq: Right Cluster Classification}
		C_{ I_{ 2 ( n + m ) } }
		\ket* { a_{ \mathbf{ p }_{ h } } }
		=
		C_{ I_{ 2 n } }
		\ket* { a_{ \mathbf{ p }_{ f_{ i } } } }
		\otimes
		C_{ I_{ 2 m } }
		\ket* { a_{ \mathbf{ p }_{ g_{ j } } } }
		=
		\ket* { \mathbf{ i } }
		\otimes
		\left(
		2^{ - m }
		\sum_{ \mathbf{ z } \in \mathbb{ B }^{ 2 m } }
		( - 1 )^{ \mathbf{ z } \cdot \mathbf{ i } }
		\ket{ \mathbf{ z } }
		\right)
		\ .
	\end{align}
	Therefore, upon measurement, the input register $IR$ will contain $\mathbf{ i } \mathbf{ x }$. Now the $2 n$ most significant qubits will contain $\mathbf{ i }$, correctly identifying the $f_{ i }$ function belonging to $F_{ 2 n }$, and the $2 m$ least significant qubits will contain a random basis ket from $\mathbb{ B }^{ 2 m }$.
\end{itemize}

\begin{figure}[htp]
	\centering
	\begin{tikzpicture} [ scale = 0.900 ] 
		\begin{yquant}[ operator/separation = 3.000 mm, register/separation = 3.000 mm, every nobit output/.style = { } ]
			qubit { $IR_{ 0 } \colon \ket{ 0 }$ } IR;
			qubit { $IR_{ 1 } \colon \ket{ 0 }$ } IR [ + 1 ];
			qubit { $\vdots$ \hspace{ 0.450 cm } } IR [ + 1 ]; discard IR [ 2 ];
			qubit { $IR_{ 2 n - 2 } \colon \ket{ 0 }$ } IR [ + 1 ];
			qubit { $IR_{ 2 n - 1 } \colon \ket{ 0 }$ } IR [ + 1 ];
			nobit AUX_m;
			qubit { $IR_{ 2 n } \colon \ket{ 0 }$ } IR [ + 1 ];
			qubit { $IR_{ 2 n + 1 } \colon \ket{ 0 }$ } IR [ + 1 ];
			qubit { $\vdots$ \hspace{ 0.450 cm } } IR [ + 1 ]; discard IR [ 7 ];
			qubit { $IR_{ 2 ( n + m ) - 2 } \colon \ket{ 0 }$ } IR [ + 1 ];
			qubit { $IR_{ 2 ( n + m ) - 1 } \colon \ket{ 0 }$ } IR [ + 1 ];
			qubit { $OR \colon \ket{ - }$ } OR;
			[
			name = Input,
			WordBlueDarker,
			line width = 0.250 mm,
			]
			barrier ( - ) ;
			[ draw = WordBlueDarker, fill = WordBlueDarker, radius = 0.400 cm ] box {\color{white} \Large \sf{H}} IR [ 0 ];
			[ draw = WordBlueDarker, fill = WordBlueDarker, radius = 0.400 cm ] box {\color{white} \Large \sf{H}} IR [ 1 ];
			[ draw = WordBlueDarker, fill = WordBlueDarker, radius = 0.400 cm ] box {\color{white} \Large \sf{H}} IR [ 3 ];
			[ draw = WordBlueDarker, fill = WordBlueDarker, radius = 0.400 cm ] box {\color{white} \Large \sf{H}} IR [ 4 ];
			[ draw = WordBlueDarker, fill = WordBlueDarker, radius = 0.400 cm ] box {\color{white} \Large \sf{H}} IR [ 5 ];
			[ draw = WordBlueDarker, fill = WordBlueDarker, radius = 0.400 cm ] box {\color{white} \Large \sf{H}} IR [ 6 ];
			[ draw = WordBlueDarker, fill = WordBlueDarker, radius = 0.400 cm ] box {\color{white} \Large \sf{H}} IR [ 8 ];
			[ draw = WordBlueDarker, fill = WordBlueDarker, radius = 0.400 cm ] box {\color{white} \Large \sf{H}} IR [ 9 ];
			[
			name = Expansion,
			WordBlueDarker,
			line width = 0.250 mm,
			]
			barrier ( - ) ;
			[ draw = RedPurple, fill = RedPurple, x radius = 0.900 cm, y radius = 0.450 cm ] box { \color{white} \Large \sf{U}$_{ g_{ j } \star f_{ i } }$} ( IR - OR );
			[
			name = Oracle,
			WordBlueDarker,
			line width = 0.250 mm,
			]
			barrier ( - ) ;
			[ draw = GreenLighter2, fill = GreenLighter2, x radius = 0.700 cm, y radius = 0.350 cm ] box { \color{white} \Large \sf{C}$_{ I_{ 2 } }$}  ( IR [ 0 ] - IR [ 1 ] );
			[ draw = GreenLighter2, fill = GreenLighter2, x radius = 0.700 cm, y radius = 0.350 cm ] box { \color{white} \Large \sf{C}$_{ I_{ 2 } }$}  ( IR [ 3 ] - IR [ 4 ] );
			[ draw = GreenLighter2, fill = GreenLighter2, x radius = 0.700 cm, y radius = 0.350 cm ] box { \color{white} \Large \sf{C}$_{ I_{ 2 } }$}  ( IR [ 5 ] - IR [ 6 ] );
			[ draw = GreenLighter2, fill = GreenLighter2, x radius = 0.700 cm, y radius = 0.350 cm ] box { \color{white} \Large \sf{C}$_{ I_{ 2 } }$}  ( IR [ 8 ] - IR [ 9 ] );
			[
			name = Classifier,
			WordBlueDarker,
			line width = 0.250 mm,
			]
			barrier ( - ) ;
			[ line width = .350 mm, draw = white, fill = black, radius = 0.400 cm ] measure IR [ 0 ];
			[ line width = .350 mm, draw = white, fill = black, radius = 0.400 cm ] measure IR [ 1 ];
			[ line width = .350 mm, draw = white, fill = black, radius = 0.400 cm ] measure IR [ 3 ];
			[ line width = .350 mm, draw = white, fill = black, radius = 0.400 cm ] measure IR [ 4 ];
			[ line width = .350 mm, draw = white, fill = black, radius = 0.400 cm ] measure IR [ 5 ];
			[ line width = .350 mm, draw = white, fill = black, radius = 0.400 cm ] measure IR [ 6 ];
			[ line width = .350 mm, draw = white, fill = black, radius = 0.400 cm ] measure IR [ 8 ];
			[ line width = .350 mm, draw = white, fill = black, radius = 0.400 cm ] measure IR [ 9 ];
			[
			name = Measurement,
			WordBlueDarker,
			line width = 0.250 mm,
			]
			barrier ( - ) ;
			output { $\ket{ \mathbf{ i } }$ } ( IR [ 0 ] - IR [ 4 ] );
			output { $\ket{ \mathbf{ x } }$ } ( IR [ 5 ] - IR [ 9 ] );
		\end{yquant}
		\scoped [ on background layer ]
		\draw
		[ LightSkyBlue1, -, >=stealth, dashed, line width = 0.750 mm ]
		( $ (Input) + ( - 10.000 mm, 4.000 mm ) $ ) node () {} -- ( $ (Measurement) + ( 10.000 mm, 4.000 mm ) $ ) node () {};
	\end{tikzpicture}
	\caption{This figure shows the operation of the BFPQC circuit for the function $g_{ j } \star f_{ i }$.}
	\label{fig: The BFPQC Quantum Circuit for $g_{ j } star f_{ i }$}
\end{figure}
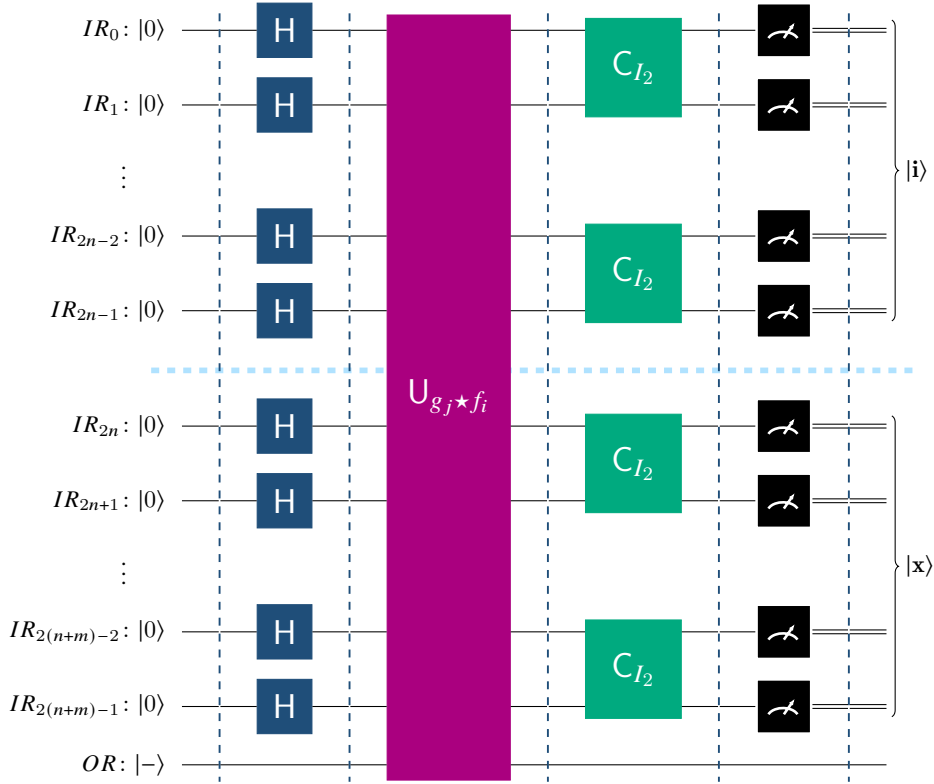
\begin{figure}[htp]
	\centering
	\begin{tikzpicture} [ scale = 0.900 ] 
		\begin{yquant}[ operator/separation = 3.000 mm, register/separation = 3.000 mm, every nobit output/.style = { } ]
			qubit { $IR_{ 0 } \colon \ket{ 0 }$ } IR;
			qubit { $IR_{ 1 } \colon \ket{ 0 }$ } IR [ + 1 ];
			qubit { $\vdots$ \hspace{ 0.450 cm } } IR [ + 1 ]; discard IR [ 2 ];
			qubit { $IR_{ 2 m - 2 } \colon \ket{ 0 }$ } IR [ + 1 ];
			qubit { $IR_{ 2 m - 1 } \colon \ket{ 0 }$ } IR [ + 1 ];
			nobit AUX_m;
			qubit { $IR_{ 2 m } \colon \ket{ 0 }$ } IR [ + 1 ];
			qubit { $IR_{ 2 m + 1 } \colon \ket{ 0 }$ } IR [ + 1 ];
			qubit { $\vdots$ \hspace{ 0.450 cm } } IR [ + 1 ]; discard IR [ 7 ];
			qubit { $IR_{ 2 ( n + m ) - 2 } \colon \ket{ 0 }$ } IR [ + 1 ];
			qubit { $IR_{ 2 ( n + m ) - 1 } \colon \ket{ 0 }$ } IR [ + 1 ];
			qubit { $OR \colon \ket{ - }$ } OR;
			[
			name = Input,
			WordBlueDarker,
			line width = 0.250 mm,
			]
			barrier ( - ) ;
			[ draw = WordBlueDarker, fill = WordBlueDarker, radius = 0.400 cm ] box {\color{white} \Large \sf{H}} IR [ 0 ];
			[ draw = WordBlueDarker, fill = WordBlueDarker, radius = 0.400 cm ] box {\color{white} \Large \sf{H}} IR [ 1 ];
			[ draw = WordBlueDarker, fill = WordBlueDarker, radius = 0.400 cm ] box {\color{white} \Large \sf{H}} IR [ 3 ];
			[ draw = WordBlueDarker, fill = WordBlueDarker, radius = 0.400 cm ] box {\color{white} \Large \sf{H}} IR [ 4 ];
			[ draw = WordBlueDarker, fill = WordBlueDarker, radius = 0.400 cm ] box {\color{white} \Large \sf{H}} IR [ 5 ];
			[ draw = WordBlueDarker, fill = WordBlueDarker, radius = 0.400 cm ] box {\color{white} \Large \sf{H}} IR [ 6 ];
			[ draw = WordBlueDarker, fill = WordBlueDarker, radius = 0.400 cm ] box {\color{white} \Large \sf{H}} IR [ 8 ];
			[ draw = WordBlueDarker, fill = WordBlueDarker, radius = 0.400 cm ] box {\color{white} \Large \sf{H}} IR [ 9 ];
			[
			name = Expansion,
			WordBlueDarker,
			line width = 0.250 mm,
			]
			barrier ( - ) ;
			[ draw = RedPurple, fill = RedPurple, x radius = 0.900 cm, y radius = 0.450 cm ] box { \color{white} \Large \sf{U}$_{ f_{ i } \star g_{ j } }$} ( IR - OR );
			[
			name = Oracle,
			WordBlueDarker,
			line width = 0.250 mm,
			]
			barrier ( - ) ;
			[ draw = GreenLighter2, fill = GreenLighter2, x radius = 0.700 cm, y radius = 0.350 cm ] box { \color{white} \Large \sf{C}$_{ I_{ 2 } }$}  ( IR [ 0 ] - IR [ 1 ] );
			[ draw = GreenLighter2, fill = GreenLighter2, x radius = 0.700 cm, y radius = 0.350 cm ] box { \color{white} \Large \sf{C}$_{ I_{ 2 } }$}  ( IR [ 3 ] - IR [ 4 ] );
			[ draw = GreenLighter2, fill = GreenLighter2, x radius = 0.700 cm, y radius = 0.350 cm ] box { \color{white} \Large \sf{C}$_{ I_{ 2 } }$}  ( IR [ 5 ] - IR [ 6 ] );
			[ draw = GreenLighter2, fill = GreenLighter2, x radius = 0.700 cm, y radius = 0.350 cm ] box { \color{white} \Large \sf{C}$_{ I_{ 2 } }$}  ( IR [ 8 ] - IR [ 9 ] );
			[
			name = Classifier,
			WordBlueDarker,
			line width = 0.250 mm,
			]
			barrier ( - ) ;
			[ line width = .350 mm, draw = white, fill = black, radius = 0.400 cm ] measure IR [ 0 ];
			[ line width = .350 mm, draw = white, fill = black, radius = 0.400 cm ] measure IR [ 1 ];
			[ line width = .350 mm, draw = white, fill = black, radius = 0.400 cm ] measure IR [ 3 ];
			[ line width = .350 mm, draw = white, fill = black, radius = 0.400 cm ] measure IR [ 4 ];
			[ line width = .350 mm, draw = white, fill = black, radius = 0.400 cm ] measure IR [ 5 ];
			[ line width = .350 mm, draw = white, fill = black, radius = 0.400 cm ] measure IR [ 6 ];
			[ line width = .350 mm, draw = white, fill = black, radius = 0.400 cm ] measure IR [ 8 ];
			[ line width = .350 mm, draw = white, fill = black, radius = 0.400 cm ] measure IR [ 9 ];
			[
			name = Measurement,
			WordBlueDarker,
			line width = 0.250 mm,
			]
			barrier ( - ) ;
			output { $\ket{ \mathbf{ x } }$ } ( IR [ 0 ] - IR [ 4 ] );
			output { $\ket{ \mathbf{ i } }$ } ( IR [ 5 ] - IR [ 9 ] );
		\end{yquant}
		\scoped [ on background layer ]
		\draw
		[ LightSkyBlue1, -, >=stealth, dashed, line width = 0.750 mm ]
		( $ (Input) + ( - 10.000 mm, 4.000 mm ) $ ) node () {} -- ( $ (Measurement) + ( 10.000 mm, 4.000 mm ) $ ) node () {};
	\end{tikzpicture}
	\caption{The above figure depicts the operation of the BFPQC circuit for the function $f_{ i } \star g_{ j }$.}
	\label{fig: The BFPQC Quantum Circuit for $f_{ i } star g_{ j }$}
\end{figure}
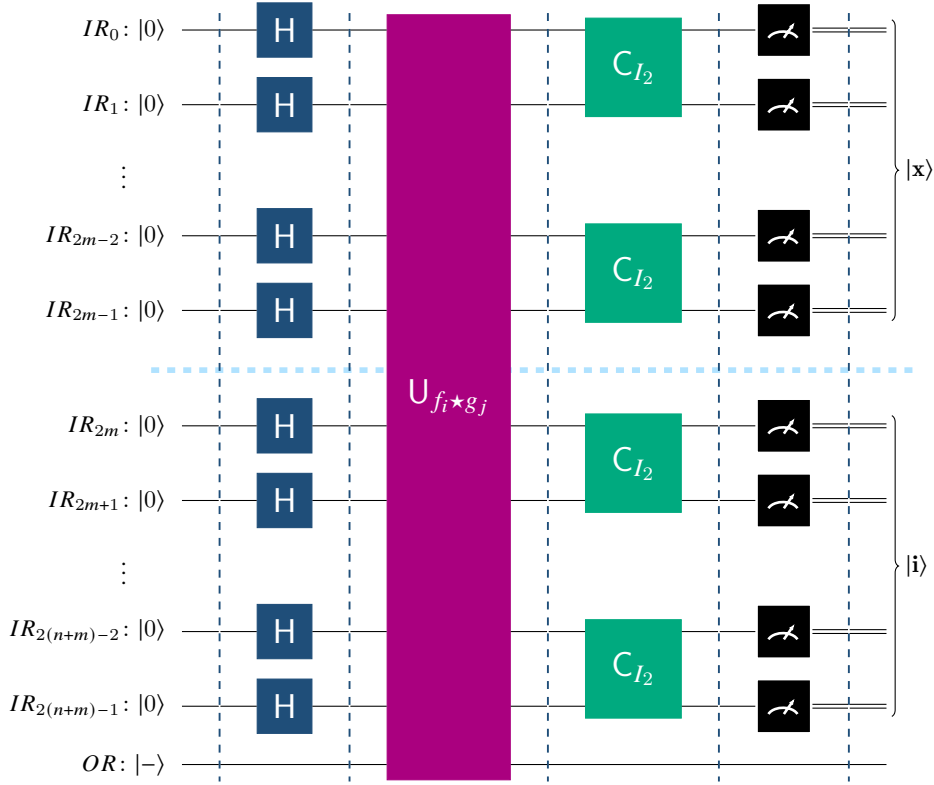

\begin{example} {Functions Outside the Promised Class} { Functions Outside The Promised Class}

Let us now consider two rounds of the Classification Game played by Alice and Bob.

\begin{itemize}
	\item	
	In the first round, Bob announces that he has chosen a function from the left cluster $G_{ 2 } \star F_{ 2 }$. This is the only information known to Alice. Then, he constructs the oracle for the secret function $g_{ 3 } \star f_{ 3 }$, where $g_{ 3 }$ is the function of $G_{ 2 }$ corresponding to the pattern bit vector $\mathbf{ e }_{ 3 }$ from \eqref{eq: Pattern Basis $B_{ 2 }$}, and $f_{ 3 }$ is from $F_{ 2 }$ and corresponds to $\mathbf{ p }_{ 3 }$ from \eqref{eq: Pattern Basis I_2}. In this scenario, Alice must use the abstract quantum circuit shown in Figure \ref{fig: The BFPQC Quantum Circuit for $g_{ j } star f_{ i }$}.
	By applying \eqref{eq: Left Cluster Classification} we see that the state of the input register $IR$ prior to the final measurement is
	\begin{align*}
		\left(
		\frac { 1 } { 2 }
		\sum_{ \mathbf{ z } \in \mathbb{ B }^{ 2 } }
		( - 1 )^{ \mathbf{ z } \cdot 11 }
		\ket{ \mathbf{ z } }
		\right)
		\ket* { 11 }
		\ .
	\end{align*}
	Figure \ref{fig:_Phase4_Histogram_StatevectorSampler___0110x1000___} shows the state of the input register $IR$ prior to measurement. Upon measurement, the input register $IR$ will be $\mathbf{ x } 11$, for some random $\mathbf{ x } \in \mathbb{ B }^{ 2 }$. In other words, the $2$ most significant qubits will contain a random basis ket from $\mathbb{ B }^{ 2 }$, and the $2$ least significant qubits will contain $11$, which reveals the $f_{ 3 }$ function. Therefore, Alice will correctly find the second component of $g_{ 3 } \star f_{ 3 }$ belonging to $F_{ 2 }$, and win this round.
\end{itemize}

\begin{itemize}
	\item	
	For the second round, Bob announces to Alice that he has picked a function from the right cluster $F_{ 2 } \star G_{ 2 }$. Accordingly, he constructs the oracle for the unknown function $f_{ 3 } \star g_{ 3 }$, where  $f_{ 3 }$ is from $F_{ 2 }$ and corresponds to $\mathbf{ p }_{ 3 }$ from \eqref{eq: Pattern Basis I_2}, and $g_{ 3 }$ from $G_{ 2 }$ corresponding to the pattern bit vector $\mathbf{ e }_{ 3 }$ from \eqref{eq: Pattern Basis $B_{ 2 }$}. For this case, Alice must use the abstract quantum circuit shown in Figure \ref{fig: The BFPQC Quantum Circuit for $f_{ i } star g_{ j }$}.
	By applying \eqref{eq: Right Cluster Classification} we conclude that the state of the input register $IR$ prior to the final measurement is
	\begin{align*}
		\ket* { 11 }
		\left(
		\frac { 1 } { 2 }
		\sum_{ \mathbf{ z } \in \mathbb{ B }^{ 2 } }
		( - 1 )^{ \mathbf{ z } \cdot 11 }
		\ket{ \mathbf{ z } }
		\right)
		\ .
	\end{align*}
	Figure \ref{fig:_Phase4_Histogram_StatevectorSampler___1000x0110___} shows the state of the input register $IR$ prior to measurement. Upon measurement, the input register $IR$ will be $11 \mathbf{ x }$, for some random $\mathbf{ x } \in \mathbb{ B }^{ 2 }$. In other words, the $2$ most significant qubits will contain $11$, while the $2$ least significant qubits will contain a random basis ket from $\mathbb{ B }^{ 2 }$. Thus, again Alice will correctly identify the first component of $f_{ 3 } \star g_{ 3 }$ belonging to $F_{ 2 }$, and win this round too.
\end{itemize}

\begin{tcolorbox}
	[
		enhanced,
		breakable,
		center title,
		fonttitle = \bfseries,
		grow to left by = 1.000 cm,
		grow to right by = 0.000 cm,
		colback = white,						
		enhanced jigsaw,						
		sharp corners,
		toprule = 0.001 pt,
		bottomrule = 0.001 pt,
		leftrule = 0.001 pt,
		rightrule = 0.001 pt,
	]
	\begin{figure}[H]
		\centering
		\includegraphics [ scale = 0.650, trim = {0.000cm 0.000cm 0.000cm 0.000cm}, clip ] {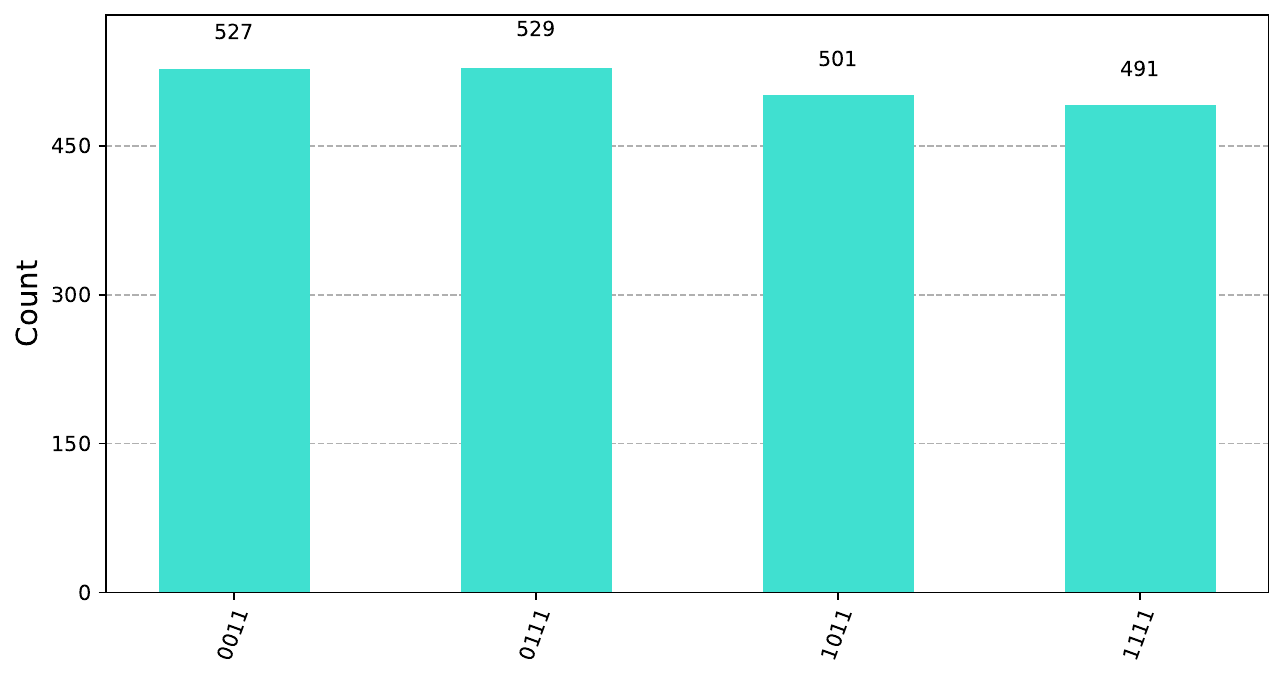}
		\caption{This figure depicts the state of the input register $IR$ prior to measurement when the unknown function is $g_{ 3 } \star f_{ 3 }$.}
		\label{fig:_Phase4_Histogram_StatevectorSampler___0110x1000___}
	\end{figure}
\end{tcolorbox}

\begin{tcolorbox}
	[
		enhanced,
		breakable,
		center title,
		fonttitle = \bfseries,
		grow to left by = 1.000 cm,
		grow to right by = 0.000 cm,
		colback = white,						
		enhanced jigsaw,						
		sharp corners,
		toprule = 0.001 pt,
		bottomrule = 0.001 pt,
		leftrule = 0.001 pt,
		rightrule = 0.001 pt,
	]
	\begin{figure}[H]
		\centering
		\includegraphics [ scale = 0.650, trim = {0.000cm 0.000cm 0.000cm 0.000cm}, clip ] {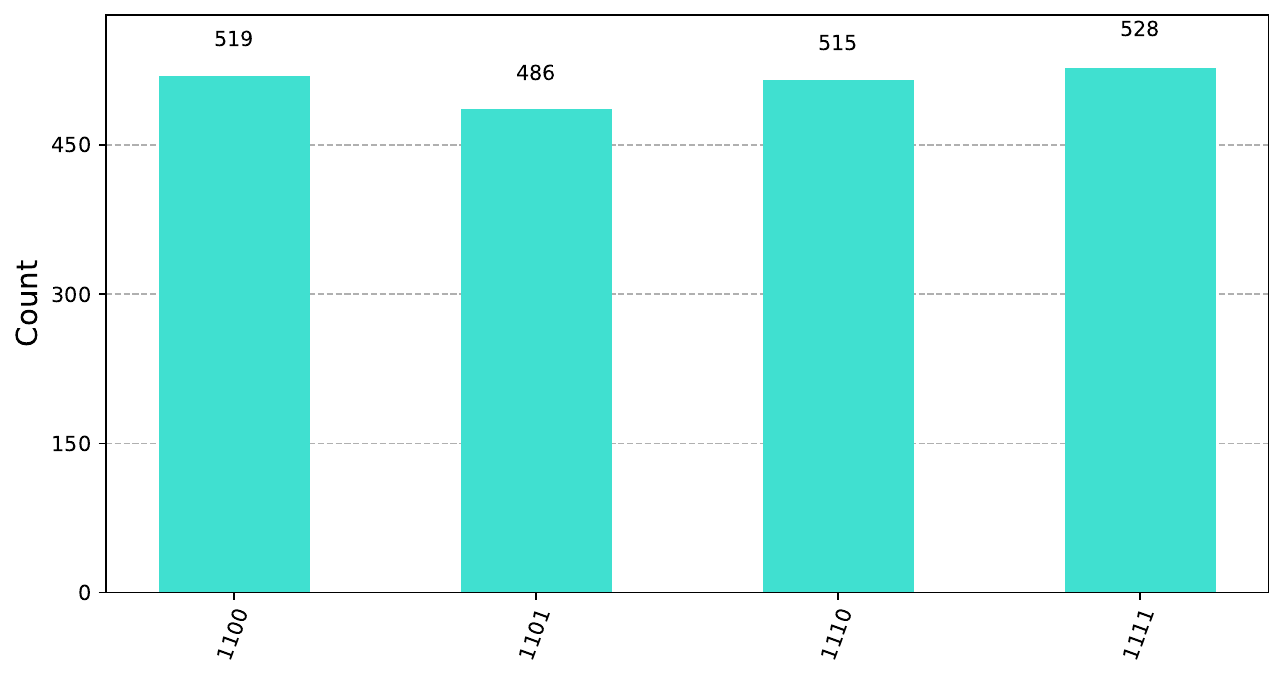}
		\caption{This figure depicts the state of the input register $IR$ prior to measurement when the unknown function is $f_{ 3 } \star g_{ 3 }$.}
		\label{fig:_Phase4_Histogram_StatevectorSampler___1000x0110___}
	\end{figure}
\end{tcolorbox}

\end{example}

\section{What did Alice learn?} \label{sec: What Did Alice Learn?}

In this Section we take a closer look at the precise amount of information that Alice can surmise from the Classification Game depending on whether Bob picks a function $h$ from the left cluster $G_{ 2 m } \star F_{ 2 n }$ or from the right cluster $F_{ 2 n } \star G_{ 2 m }$. To clearly understand the details, it will be expedient to recall Proposition \ref{prp: Equivalence Of Products} concerning the equivalence of the different types of products, and focus on the pattern bit vectors. So, let us write down the three pattern bit vectors $\mathbf{ p }_{ f_{ i } }$, $\mathbf{ p }_{ g_{ j } }$, and $\mathbf{ p }_{ h }$ in their expanded form

\begin{align}
	\label{eq: Pattern Bit Vectors Expanded Form}
	\left
	\{
	\
	\begin{aligned}
		\mathbf{ p }_{ f_{ i } }
		&=
		a_{ 2^{ 2 n } - 1 }
		\dots
		a_{ 1 }
		a_{ 0 }
		\\
		\mathbf{ p }_{ g_{ j } }
		&=
		b_{ 2^{ 2 m } - 1 }
		\dots
		b_{ 1 }
		b_{ 0 }
		\\
		\mathbf{ p }_{ h }
		&=
		c_{ 2^{ 2 ( n + m ) } - 1 }
		\dots
		c_{ 1 }
		c_{ 0 }
	\end{aligned}
	\
	\right
	\}
	\ .
\end{align}

Although the amount of information is the same in both cases, there some important subtle differences. For this reason it is instructive to distinguish between the following two cases based on whether the unknown function belongs to the left or right cluster.

First, we examine the case where $h = g_{ j } \star f_{ i }$. Then, by Proposition \ref{prp: Equivalence Of Products} we know that

\begin{align}
	\label{eq: Left Cluster Product Form}
	\mathbf{ p }_{ h }
	&=
	\mathbf{ p }_{ g_{ j } }
	\odot
	\mathbf{ p }_{ f_{ i } }
	\Leftrightarrow
	\nonumber
	\\
	c_{ 2^{ 2 ( n + m ) } - 1 }
	\dots
	c_{ 1 }
	c_{ 0 }
	&=
	\underbrace { \colorbox {RedPurple!48} { $a_{ 2^{ 2 n } - 1 }^{ 2^{ 2 m } - 1 } \dots a_{ 1 }^{ 2^{ 2 m } - 1 } a_{ 0 }^{ 2^{ 2 m } - 1 }$ } }_{ \text{ block } 2^{ 2 m } - 1 }
	\cdots
	\underbrace { \colorbox {RedPurple!24} { $a_{ 2^{ 2 n } - 1 }^{ 1 } \dots a_{ 1 }^{ 1 } a_{ 0 }^{ 1 }$ } }_{ \text{ block } 1 }
	\
	\underbrace { \colorbox {RedPurple!12} { $a_{ 2^{ 2 n } - 1 }^{ 0 } \dots a_{ 1 }^{ 0 } a_{ 0 }^{ 0 }$ } }_{ \text{ block } 0 }
	\ ,
\end{align}

where

\begin{align}
	\label{eq: Left Cluster Block Form}
	a_{ 2^{ 2 n } - 1 }^{ r } \dots a_{ 1 }^{ r } a_{ 0 }^{ r }
	=
	\begin{cases}
		\ a_{ 2^{ 2 n } - 1 } \dots a_{ 1 } a_{ 0 }					& \text{if } b_{ r } = 0 
		\\[ 1.000 ex ]
		\ \overline{ a_{ 2^{ 2 n } - 1 } \dots a_{ 1 } a_{ 0 } }	& \text{if } b_{ r } = 1 
	\end{cases}
	\ ,
	\quad
	0 \leq r \leq 2^{ 2 m } - 1
	\ .
\end{align}

In view of the above analysis, we may summarize the following facts.

\begin{enumerate}
	[ left = 1.300 cm, labelsep = 0.500 cm, start = 1 ]
	\renewcommand \labelenumi { (\textbf{KL}$_{ \theenumi }$) }
	\item	Alice knows that the pattern bit vector of the $h$ function, which encodes its behavior, consists of $2^{ 2 m }$ blocks, each of length $2^{ 2 n }$. Alice knows $f_{ i }$ and, therefore, knows that every block is either equal to $\mathbf{ p }_{ f_{ i } }$ or its negation.
	\item	The modulo $2$ sum of any two distinct bits $0 \leq k < l \leq 2^{ 2 n } - 1$ of the same block is equal to the  modulo $2$ sum of the corresponding bits in every other block:
	\begin{align}
		\label{eq: Left Cluster Bit Correlations}
		\underbrace { \colorbox {Azure2} { $a_{ l }^{ 2^{ 2 m } - 1 } \oplus a_{ k }^{ 2^{ 2 m } - 1 }$ } }_{ \text{ block } 2^{ 2 m } - 1 }
		=
		\cdots
		=
		\underbrace { \colorbox {Azure2} { $a_{ l }^{ 1 } \oplus a_{ k }^{ 1 }$ } }_{ \text{ block } 1 }
		=
		\underbrace { \colorbox {Azure2} { $a_{ l }^{ 0 } \oplus a_{ k }^{ 0 }$ } }_{ \text{ block } 0 }
	\end{align}
	\item	It is worth noting that knowledge of a single additional bit in any block automatically reveals to Alice the entire block.
	\item	However, this information is not enough to allow Alice to reconstruct $\mathbf{ p }_{ h }$.
\end{enumerate}

Now, we examine the case where $h = f_{ i } \star g_{ j }$. This time Proposition \ref{prp: Equivalence Of Products} implies

\begin{align}
	\label{eq: Right Cluster Product Form}
	\mathbf{ p }_{ h }
	&=
	\mathbf{ p }_{ f_{ i } }
	\odot
	\mathbf{ p }_{ g_{ j } }
	\Leftrightarrow
	\nonumber
	\\
	c_{ 2^{ 2 ( n + m ) } - 1 }
	\dots
	c_{ 1 }
	c_{ 0 }
	&=
	\underbrace { \colorbox {GreenLighter2!48} { $b_{ 2^{ 2 m } - 1 }^{ 2^{ 2 n } - 1 } \dots b_{ 1 }^{ 2^{ 2 n } - 1 } b_{ 0 }^{ 2^{ 2 n } - 1 }$ } }_{ \text{ block } 2^{ 2 n } - 1 }
	\cdots
	\underbrace { \colorbox {GreenLighter2!24} { $b_{ 2^{ 2 m } - 1 }^{ 1 } \dots b_{ 1 }^{ 1 } b_{ 0 }^{ 1 }$ } }_{ \text{ block } 1 }
	\
	\underbrace { \colorbox {GreenLighter2!12} { $b_{ 2^{ 2 m } - 1 }^{ 0 } \dots b_{ 1 }^{ 0 } b_{ 0 }^{ 0 }$ } }_{ \text{ block } 0 }
	\ ,
\end{align}

where

\begin{align}
	\label{eq: Right Cluster Block Form}
	b_{ 2^{ 2 m } - 1 }^{ r } \dots b_{ 1 }^{ r } b_{ 0 }^{ r }
	=
	\begin{cases}
		\ b_{ 2^{ 2 m } - 1 } \dots b_{ 1 } b_{ 0 }					& \text{if } a_{ r } = 0 
		\\[ 1.000 ex ]
		\ \overline{ b_{ 2^{ 2 m } - 1 } \dots b_{ 1 } b_{ 0 } }	& \text{if } a_{ r } = 1 
	\end{cases}
	\ ,
	\quad
	0 \leq r \leq 2^{ 2 n } - 1
	\ .
\end{align}

In view of the above analysis, we may summarize the following facts.

\begin{enumerate}
	[ left = 1.400 cm, labelsep = 0.500 cm, start = 1 ]
	\renewcommand \labelenumi { (\textbf{KR}$_{ \theenumi }$) }
	\item	Alice knows that the pattern bit vector of the $h$ function, which encodes its behavior, consists of $2^{ 2 n }$ blocks, each of length $2^{ 2 m }$. Now, the fact that Alice knows $f_{ i }$ implies that Alice knows with certainty which block is equal to $\mathbf{ p }_{ g_{ j } }$ and which block is equal to its negation.
	\item	For any position $k$, $0 \leq k \leq 2^{ 2 m } - 1$, Alice knows in which blocks the bit in the same position $k$ has the same value and in which blocks has the opposite value.
	\item	In this case, knowledge of a single additional bit in any block automatically reveals the corresponding bit to all other blocks.
	\item	Unfortunately, this information is also not enough to allow Alice to reconstruct $\mathbf{ p }_{ h }$.
\end{enumerate}

\section{Discussion and conclusions} \label{sec: Discussion and Conclusions}

Advanced research on the classification of Boolean functions that satisfy specific properties, i.e., belong to a particular class, is abundant in the literature. Yet, to the best of our knowledge, no earlier studies have examined what information might be revealed when the input Boolean function does not fit into this category. This paper offers a fresh research perspective that demonstrates that significant knowledge can still be obtained under appropriate circumstances.

In this paper, we have studied study when a specific algorithm is applied on functions do not belong to the promised class, but are still relatively close to its elements. For this purpose we have introduced a novel concept that defines ``nearness'' between Boolean functions. We illustrate this concept by showing that, as long as the input function is close enough to the promised class, the classification algorithm can still yield insightful information about its behavioral pattern. To the best of our knowledge, this study is the first to demonstrate the benefits of using quantum classification algorithms to look at functions that are not part of the promised class in order to get a glimpse of important data.

\begin{tcolorbox}
	[
		enhanced,
		breakable,
		center title,
		fonttitle = \bfseries,
		colbacktitle = Turquoise4,
		coltitle = white,
		title = Synopsis,
		grow to left by = 0.000 cm,
		grow to right by = 0.000 cm,
		colframe = Turquoise4,
		colback = SeaGreen1!20,
		enhanced jigsaw,			
		sharp corners,
		boxrule = 0.500 pt,
	]
	When the unknown input function $h$ belongs to the left or right cluster of a function $f$ within the promised class, meaning that $h$ can written either as $g \star f$, or $f \star g$, where $g$, and, thus, $h$ is outside the promised class, the classification algorithm BFPQC can conclusively unveil $f$ with a single query. This furnishes important information about the behavioral pattern of $h$, albeit not the complete information.
\end{tcolorbox}
\bibliographystyle{ieeetr}
\bibliography{QCOPC}

\end{document}